\documentclass[aps, nofootinbib, twocolumn, showpacs,amsfonts,amsmath,floatfix, superscriptaddress]{revtex4-1}

\usepackage[utf8]{inputenc} % So université comes out correctly
\usepackage{graphicx}% Include figure files
\usepackage[usenames,dvipsnames,svgnames,table]{xcolor}
\usepackage[caption=false,labelfont=large]{subfig}
\graphicspath{{./images/}}
\usepackage{dcolumn}% Align table columns on decimal point
\usepackage{mathtools} % for small matrices and other tricks
\usepackage{bm}% bold math
\usepackage{empheq} % for aligned braces equations
\usepackage{changepage} % for adjustwidth
\usepackage{microtype} %typographic optimizations
\usepackage{hyperref}% add hypertext capabilities

\newcommand{\Bra}[2][]{\left<#2\right|_{#1}}
\newcommand{\Ket}[2][]{\left|#2\right>_{\hspace{-0.1em}#1}}

\newcommand{\be}{\begin{equation}}
\newcommand{\ee}{\end{equation}}
\newcommand{\ba}[1]{\begin{array}{#1}}
\newcommand{\ea}{\end{array}}
\newcommand{\bpm}[1][c]{\begin{pmatrix*}[#1]}
\newcommand{\epm}{\end{pmatrix*}}
\newcommand{\bpsm}[1][c]{\begin{psmallmatrix*}[#1]}
\newcommand{\epsm}{\end{psmallmatrix*}}
\newcommand{\bal}{\begin{align}}
\newcommand{\eal}{\end{align}}
\newcommand{\baln}{\begin{align*}}
\newcommand{\ealn}{\end{align*}}
\newcommand{\lr}[1]{\left( #1 \right)}
\newcommand{\ddd}{\mathrm{d}}

\newcommand{\bs}[1]{\boldsymbol{#1}}

\newcommand{\bxi}{\bs{\xi}}

\newcommand{\norm}[1]{\left\lVert#1\right\rVert}

\newcommand\restr[2]{{% we make the whole thing an ordinary symbol
  \left.\kern-\nulldelimiterspace % automatically resize the bar with \right
  #1 % the function
  \vphantom{\big|} % pretend it's a little taller at normal size
  \right|_{#2} % this is the delimiter
  }} %restriction of a function to subset of domain

\newcommand{\prlsection}[1]{\section{#1}}

\begin{document}
%\preprint{APS/123-QED}

\title{Random coding for sharing bosonic quantum secrets}% Force line breaks with \\

\author{Francesco Arzani}
\email{fra.arzani@gmail.com}
\affiliation{Universit\' e de Lorraine, CNRS, Inria, LORIA, F 54000 Nancy, France}
\affiliation{Sorbonne Université, CNRS, Laboratoire d’Informatique de Paris 6, F-75005 Paris, France}
\author{Giulia Ferrini}
\affiliation{Department of Microtechnology and Nanoscience (MC2), Chalmers University of Technology, SE-412 96 Gothenburg, Sweden}
\author{Fr\' ed\' eric Grosshans}
\affiliation{Laboratoire Aim\' e Cotton, CNRS, Univ.\@ Paris-Sud, ENS Cachan, Univ.\@ Paris Saclay, 91405 Orsay Cedex, France}
\affiliation{Sorbonne Université, CNRS, Laboratoire d’Informatique de Paris 6, F-75005 Paris, France}
\affiliation{Paris Center for Quantum Computing, CNRS FR3640, Paris, France}
\author{Damian Markham}
\affiliation{Sorbonne Université, CNRS, Laboratoire d’Informatique de Paris 6, F-75005 Paris, France}
\affiliation{Paris Center for Quantum Computing, CNRS FR3640, Paris, France}

\date{\today}

\begin{abstract}

We consider a protocol for sharing quantum states using continuous variable systems.
Specifically we introduce an encoding procedure where bosonic modes in arbitrary secret states are mixed with several ancillary squeezed modes through a passive interferometer.
We derive simple conditions on the interferometer for this encoding to define a secret sharing protocol and we prove that they are satisfied by almost any interferometer. 
This implies that, if the interferometer is chosen uniformly at random, the probability that it may not be used to implement a quantum secret sharing protocol is zero. 
Furthermore, we show that the decoding operation can be obtained and implemented efficiently with a Gaussian unitary using a number of single-mode squeezers that is at most twice the number of modes of the secret, regardless of the number of players.
We benchmark the quality of the reconstructed state by computing the fidelity with the secret state as a function of the input squeezing.

\end{abstract}

\maketitle
%----------------------------------------------------------------------------------------------------------------------------
\prlsection{Introduction}%
Quantum systems are notoriously fragile: 
small losses or weak interactions with the outside world usually destroy quantum coherence.
Since quantum information cannot be copied \cite{noCloning}, %
any leakage of information leads to its destruction in the original system.
To fully retrieve it, one usually needs full 
control over the environment. 
This loss of coherence is at the heart of quantum information, whether we want to fight
it \cite{shor,FTGottesman,Knill342} %
or impose it on an adversary \cite{QKD,QKDrev,money}, % 
but it plays an important role in a broader area of physics, including
thermodynamics \cite{decohThermo1,decohThermo2}, %
quantum control \cite{decohControl1,decohControl2}, and
black hole physics \cite{BHasMirrors,OMG,haydenReplication,sandersRelavitistic,sandersEfficient}.

Among the strategies devised to try and overcome this fragility are %
quantum error correcting (QEC) codes \cite{QECCcodes, hay2008ran} %
and quantum secret sharing (QSS) schemes \cite{gottesmanTheorySS, Hillery1999, cleveSS}.

In QSS schemes, a \emph{dealer} delocalizes the information between several
players, so that authorized subsets of them (\emph{access parties})
can fully reconstruct the original information without the shares of 
the other players. Unauthorized sets (\emph{adversaries}) on the other hand get in principle no information about the secret. QSS schemes are equivalent to erasure correcting codes 
\cite{gottesmanTheorySS}, protecting against loss of part of the system. 
As well as protecting information, they have many applications in quantum information, such as secure multiparty computation \cite{ben2006secure}. 
Most QSS and QEC schemes \cite{Hillery1999, cleveSS, QECCcodes}
are highly structured. %
However, random codes have been proven to optimally protect the state of a set of qubits 
from erasure errors \cite{hay2008ran}. Furthermore, their randomness makes them a natural
model in a variety of physical scenarios where information is lost.  

Most of these results are for two-dimensional, qubit encodings. Alternative to qubits, information can be encoded in the state of infinite-dimensional quantum systems, known as continuous-variable (CV) systems. CV systems are of great practical importance in quantum technologies \cite{gaussiInfoRev}: the possibility to experimentally generate entanglement in a deterministic fashion makes them interesting candidates for the realization of quantum communication and computation protocols. Several CV generalizations  of QSS \cite{tyc1,tyc2,MvLSS,weedbrookSS} and erasure-correcting codes \cite{cerfECC, vLoockECC} have been proposed, and some have  been experimentally demonstrated \cite{lance2004tripartite,Armstrong2015,cai2017multimode}. Each of these schemes, however, requires encoding the secret in carefully chosen states. No CV random code has been proposed to date. This gap poses serious limitations to the experimental realization of CV-QSS. For example, unless the experimental setup is specifically tailored for the task, CV-QSS could not be carried out, or experimental imperfections might hinder its implementation. As in the qubit case, one may also expect applications of random coding beyond QSS and quantum information \cite{BHasMirrors,OMG,haydenReplication,sandersRelavitistic,sandersEfficient}.

In this work, we fill this gap by introducing a form of random coding for CV. Namely, we show that QSS can be implemented in bosonic systems mixing a secret state with squeezed states, the workhorse of CV quantum information \cite{CVQIvLoockBraunstein,ferraro2005gaussian}, through \emph{almost any} energy preserving transformation. The latter correspond to passive interferometers in the optical setting. Our approach also generalizes earlier proposals by allowing the secret to be an arbitrary multimode state, as long as enough players are considered. We show that for almost any passive transformation there exists a decoding scheme, that each authorized set can construct efficiently, such that the secret can be recovered to arbitrary precision, provided the initial squeezing is high enough. The decoding only requires Gaussian resources, considered relatively easy to implement experimentally \cite{bachor2004guide,ferraro2005gaussian}. We show that in the optical case, decoding can be achieved by interferometry, homodyne detection and a fixed number of single-mode squeezers. We stress that our results follow from simple linear algebra and general considerations on the number of modes.

These results have immediate experimental and technological applications. Indeed, they imply that almost any experimental setup involving squeezed states can be used for QSS. Moreover, small deviations of the setup from a theoretical target one are not important, as long as they can be characterized. This opens the possibility to share resource states securely over a network of CV systems with arbitrarily distributed entanglement links, which may pave the way to server-client architectures for CV-quantum computation. But the relevance of CV random codes is not limited to their practicality. Optimality of random erasure correcting codes for qubits was used in a seminal article to estimate the rate of information leakage from black holes through Hawking radiation \cite{BHasMirrors}. The most relevant objects in this setting are however fields, namely CV systems. This stimulated work applying CV techniques, notably related to QSS, in relativistic contexts \cite{OMG,haydenReplication,sandersRelavitistic,sandersEfficient}. The existence of efficient CV random QSS codes may open new avenues for tackling the black-hole information puzzle and related fundamental questions.

The remainder of the article is structured as follows. Some background information is recalled in Sec.~\ref{sec:back}. In Sec.~\ref{sec:GenSSSec} we describe the encoding procedure the dealer uses to share the secret with all players. In Sec.~\ref{sec:decCond} we derive conditions ensuring that any sufficiently large group of players can retrieve the secret state. The decoding operations that the players can carry out when these conditions are satisfied are further described in Sec.~\ref{sec:preciseDec}. A precise formulation of our main result is given in Sec.~\ref{sec:haar} together with a sketch of the proof. In an ideal setting, access parties should get full information about the secret and adversaries should get none. As it is often the case in CV, the ideal situation is never achieved in physical scenarios where only finite squeezing is available. Sec.~\ref{sec:stuff} and Sec.~\ref{sec:unsets} discuss the amount of information retrieved by the authorized players and the adversaries, respectively, in the finite squeezing scenario. Final remarks in Sec.~\ref{sec:conclusions} conclude the article. More details about the derivations can be found in the Appendices.
 
\prlsection{CV quantum optics\label{sec:back}} 
A convenient way to study an $n$ mode bosonic system is through the $2n$-dimensional 
phase space. The $2n$ components of the quadrature vector $\bs{\xi} = \lr{ \bs{q}^T, \bs{p}^T }^T$ are the position and momentum operators, obeying the canonical commutation relations \begin{equation}\left[ \xi_j, \xi_l \right] = i J^{\lr{n}}_{jl},\label{eq:genComm}\end{equation}
with $J^{\lr{n}}$ the standard symplectic form % 
\begin{equation}
J^{\lr{n}} =\left( \begin{array}{cc} \bs{0}_n & \mathbb{I}_n \\ -\mathbb{I}_n & \bs{0}_n \end{array} \right),
\end{equation} 
 $\bs{0}_n$ and $\mathbb{I}_n$ being zero and identity $n\times n$ matrices.
The state of a $n$-mode system is characterized by its Wigner function
$W\lr{\bs{q} , \bs{p}}$ \footnote{%
    Here, $\bf q$ and $\bf p$ are \emph{real}-valued $n$ dimensional vectors, not vectors 
    of operators. In the following we use the same symbols for vectors of quadrature operators and phase-space variables, the meaning should be clear from the context.}, a quasi-probability distribution defined on phase-space \cite{ferraro2005gaussian}.
Gaussian states are naturally defined as those the Wigner function of which is Gaussian, 
and they are fully characterized by the mean and covariance matrix of the 
quadrature vector $\bs{\xi}$. 

Gaussian transformations---preserving the Gaussian character of the state---%
are an essential subset of physical transformations, since they can be 
implemented deterministically in quantum optics experiments with existing technologies.
Unitary Gaussian transformations are elegantly described by the formalism of symplectic matrices.
In the Heisenberg picture, the action of a unitary Gaussian operation $U_G$ can be expressed with a slight abuse of notation as a linear map\cite{ferraro2005gaussian,gaussiInfoRev}\begin{equation}
 U_G ^\dagger  \bs{\xi} U_G  = S \bs{\xi} + \bs{\eta},
\end{equation} 
where $S$ is a $2n\times 2n$ real symplectic matrix and $\bs{\eta}$ is a vector of real numbers~\cite{pramana}. 
Symplectic matrices acting on $n$ modes are the matrices $S$ preserving the 
standard symplectic form: $ S J^{\lr{n}} S^T = J^{\lr{n}}$. Under matrix multiplication, they form the group $\mathrm{Sp}\lr{2n,\mathbb{R}}$. If displacements are included, amounting to phase-space translations, one gets the so-called inhomogeneous symplectic group.
Of specific interest are squeezing and passive operations. 
Squeezing does not conserve photon number~\cite{ferraro2005gaussian} 
and is usually realized through nonlinear optical processes. 
Independent squeezing operations on each mode are represented by diagonal symplectic matrices $K=\mathrm{diag}\lr{e^{r_1},\ \ldots,\ e^{r_n},\ e^{-r_1},\ \ldots,\ e^{-r_n}}$,
where $r_i$ is the squeezing parameter of mode $i$ \cite{leonhardt1997measuring}. Passive operations are defined as photon-number preserving Gaussian unitaries
and correspond to linear optics, represented by the subgroup $L\lr{n} = \mathrm{Sp}\lr{2n,\mathbb{R}}  \cap \mathrm{O}\lr{2n}  $ of orthogonal, symplectic matrices~\cite{pramana}. 
Each $S_L \in L\lr{n}$ corresponds to a  $n\times n$ unitary matrix 
$X + i Y \in U\lr{n}$ such that
\begin{equation}
S_L = \left(\begin{array}{cc} X & -Y \\ Y & X \end{array} \right). 
\end{equation} 
This allows us to speak interchangeably of passive interferometers or the corresponding
symplectic and unitary matrices.

We recall for later convenience that given two vectors $\bs{x},\ \bs{y} \in \mathbb{R}^{2n}$, their symplectic product is defined as \begin{equation}
\langle \bs{x},\bs{y}\rangle \equiv \bs{x}^TJ^{\lr{n}}\bs{y}.
\end{equation} We denote by $\bs{x}\cdot \bs{y}$ the ordinary Euclidean product $\bs{x}\cdot \bs{y} = \sum_j x_j y_j$ and by $\norm{\bs{x}} = \sqrt{\bs{x}\cdot\bs{x}}$ the Euclidean norm. Note that, formally, taking the dot product between a vector of real numbers and the vector of quadratures results in a linear combination of quadrature operators. The commutator between two such combinations is simply related to the symplectic product of the vectors \begin{equation}
\left[\bs{x}\cdot \bs{\xi}, \bs{y}\cdot \bs{\xi} \right] = i \langle \bs{x},\bs{y}\rangle \label{eq:symplProdComm}
\end{equation} as can be checked using Eq.~(\ref{eq:genComm}). A basis $\left\{\bs{x}_j\right\}$ of $\mathbb{R}^{2n}$ such that $\langle \bs{x}_j,\bs{x}_l\rangle = J^{\lr{n}}_{jl} $ is called  \emph{symplectic basis}.

\prlsection{Encoding\label{sec:GenSSSec}}
\begin{figure*}
\centering
\includegraphics[width=0.8\textwidth]{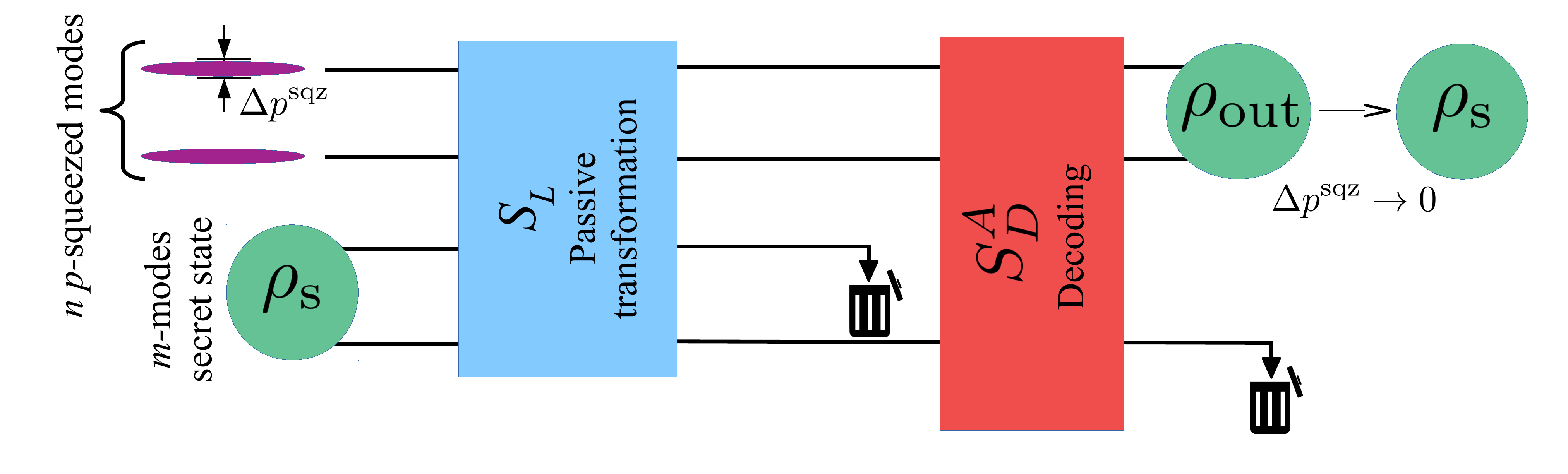}
\caption{A sketch of the encoding procedure with $n=2$, $m=2$, followed by decoding from the share of the access party consisting of the first, second and fourth mode, while the third mode, corresponding to an adversary, is discarded. As shown in the Sec.~\ref{sec:decCond}, the secret can be recovered from any $m + \lceil\frac{n}{2} \rceil =3 $ out of the four modes. The decoded state $\rho_\mathrm{out}$ converges to the secret state $\rho_\mathrm{s}$ as the input squeezing increases and the last mode can also be discarded after the decoding (see Sec.~\ref{sec:preciseDec}).}
\label{fig:encoding}
\end{figure*}
We consider the following encoding scheme (see Fig.\@~\ref{fig:encoding}): the dealer couples $m$ modes in a secret state $\rho_\mathrm{s}$  to $n$ squeezed modes in a passive interferometer described by the symplectic matrix $S_L$. We assume that each mode's momentum quadrature is squeezed ($r_i>0$). This simplifies the notation but implies no loss of generality, as local phase-space rotations aligning the squeezing directions correspond to linear optics and can be included in the interferometer.

We  denote the vector containing all input quadratures by $  \bs{\xi}^\mathrm{in} = \lr{ \lr{\bs{q}^\mathrm{sqz}}^T,  \lr{ \bs{q}^\mathrm{s}}^T, \lr{ \bs{p}^\mathrm{sqz}}^T, \lr{\bs{p}^\mathrm{s}}^T  }^T$
where the quadratures of the $j$th squeezed mode are related to the vacuum 
quadratures by $ q^\mathrm{sqz}_j =  e^{r_j}q^{\lr{0}}_j$, 
$ p^\mathrm{sqz}_j =  e^{-r_j}  p^{\lr{0}}_j $. After the interferometer the vector of quadrature operators is transformed as
\begin{equation}
  \bs{\xi}^\mathrm{net} = \left( \begin{array}{c}  
\bs{q}^\mathrm{net} \\ \bs{q}^\mathrm{d} \\ \bs{p}^\mathrm{net} \\ \bs{p}^\mathrm{d}   \end{array} \right)
      =S_L \bs{\xi}^\mathrm{in} 
      =  \left( \begin{array}{cc} X & -Y \\ Y & X  \end{array} \right) \bs{\xi}^\mathrm{in}. \label{eq:encodingFull}
\end{equation} 
One of the output modes of the interferometer is then sent to each of the players.

\prlsection{Decodability conditions \label{sec:decCond}} We now investigate the conditions that the symplectic matrix $S_L$ must satisfy and the relations between $k$, $m$ and $n$ in order for any set of $k$ or more players to be able to access the secret quadratures. Specifically, for each authorized set we look for $2m$ independent linear combinations of quadratures that do not involve the antisqueezed quadratures $q^{\mathrm{sqz}}_j$ and contain one of the secret quadratures each. We will later show that if such combinations exist, all information about the secret can be accessed by any group of $k$ or more players. We will require $k\geq m$, otherwise the players cannot reconstruct a general state of $m$ modes.

Consider a subset of players $A = \left\{a_1,\ a_2,\ \ldots,\ a_k \right\}$ who are given the modes with quadratures \begin{equation}
  \bs{\xi}^A =\bpm \bs{Q}^A \\\bs{P}^A \epm,\quad  \begin{aligned}
  \bs{Q}^A &= \bpm q_{a_1}^\mathrm{net},\ q_{a_2}^\mathrm{net} ,\ \ldots, \ q_{a_k}^\mathrm{net} \epm ^T,   \\
   \bs{P}^A &= \bpm p_{a_1}^\mathrm{net},\ p_{a_2}^\mathrm{net} ,\ \ldots, \ p_{a_k}^\mathrm{net} \epm ^T. \end{aligned}
\end{equation} $A$ need to cancel the contribution of the antisqueezed quadratures. Let us rewrite Eq.~(\ref{eq:encodingFull}) as
\begin{equation} \label{eq:simpleAP}
  \bs{\xi}^A
    = M^A \bs{q}^\mathrm{sqz} +N^A \bs{p}^\mathrm{sqz} 
      + H^A \bs{\xi}^\mathrm{s}  
\end{equation} 
where the entries of the matrices $M^A$, $N^A$, and $H^A$ are defined by the coefficients of $S_L$ and $ \bs{\xi}^\mathrm{s}$ collects the secret quadratures. Any linear combination of the $\bs{\xi}^A$s can be written  $\bs{v}^T \bs{\xi} ^A$ with $\bs{v}\in\mathbb{R}^{2k}$.  
According to Eq.\@~\eqref{eq:simpleAP}, the product $\bs{v}^T \bs{\xi} ^A$ does not contain any antisqueezed quadrature iff $\bs{v}$ lies in the kernel of $\lr{M^A}^T$. 
By construction, $M^A$ has $2k$ rows and $n$ columns, therefore \begin{equation}\label{eq:kernel}
\dim\lr{\ker(M^A)^T}\ge 2k-n
\end{equation}  
Then, if $k\geq m + \lceil\frac{n}{2}\rceil$ it is always possible to find $2m$ linearly independent vectors $\bs{v}_1,\ldots,\bs{v}_{2m}\in \ker\lr{M^A}^T$ (here $\lceil x\rceil$ denotes the smallest integer greater than or equal to $x$). 
Let us suppose this condition is satisfied and organise the vectors $\bs{v}_j$ as rows of a matrix $R$. 
Applying $R$ to $\bs{\xi}^A$ we get 
\begin{align}
  R \bs{\xi}^A &= RM^A \bs{q}^\mathrm{sqz} + RN^A \bs{p}^\mathrm{sqz} + RH ^A \bs{\xi}^\mathrm{s}   \\  & \equiv  RN^A \bs{p}^\mathrm{sqz}  + T \bs{\xi}^\mathrm{s}      \label{eq:T}
\end{align} 
where the last line defines the $2m\times 2m$ matrix $T = RH^A$.
The access party $A$ can then decode the secret iff $T$ is invertible. 
Indeed, multiplying $T^{-1}$ by Eq.\@~\eqref{eq:T} and 
defining $D \equiv T^{-1}R$, $B = T^{-1}RN^A $, leads to
\begin{equation} \label{eq:simpleDecoded}
  D\bs{\xi}^A = B\bs{p}^\mathrm{sqz} + \bs{\xi}^\mathrm{s}     .
\end{equation} 
So when $A$ measure the linear combination of quadratures defined by  the $j$th row of $D$, the outcomes will follow the same probability distribution as $\xi^\mathrm{s}_j$, apart from random displacements drawn from a Gaussian probability distribution, due to the term $B \bs{p}^\mathrm{sqz} $. 
These displacements decrease with increasing input squeezing, 
ultimately vanishing for infinite squeezing. In this limit, the access party can 
perfectly sample from the original secret state. 
Note that real linear combinations of the rows of $D$ are linear combinations of the $\xi^\mathrm{s}_j$ plus the squeezed quadratures, so $A$ can also measure arbitrary quadratures of the secret (see below). An alternative description based on Wigner functions can be found in Appendix~\ref{sec:SupplgenIn}. %

In summary, $A$ can reconstruct the secret if it is composed of at least $m + \lceil\frac{n}{2}\rceil$ players and the matrix $T$ in Eq.\@~(\ref{eq:T}) is not singular.

Given any linear optical network $S_L$, these two conditions determine the authorized subsets of players, that is the access structure. It is not necessary to construct $T$ explicitly to check whether $\det T\neq 0$ as we show in Appendix~\ref{sec:eqTinv} that this is equivalent to $\det\lr{ M^A \   H^A } \neq0$. The latter condition explicitly involves the coefficients of $S_L$, which will be useful to prove our main result.

\prlsection{Decoding \label{sec:preciseDec}} We now clarify in which sense the above conditions allow the access party to decode the secret. 
Consider an access party $A$ and suppose the conditions in the previous section are met. 
Clearly, $A$ can measure the linear combinations defined by $D\bs{\xi}^A$ 
by combining the results of local homodyne detections. 
Indeed, Eq.\@~(\ref{eq:simpleDecoded}) can be rewritten
\begin{align}
  \xi_j^\mathrm{s} + \sum_{l=1}^{n} B_{jl} p^\mathrm{sqz}_l  &=  \sum _{l=1} ^{k} \lr{D_{jl} Q^A_l + D_{j,l+k} P^A_l }\\
    & = \sum _{l=1} ^{ k} \alpha_{jl} \lr{\cos\theta_{jl} Q^A_l + \sin\theta_{jl} P^A_l }
\end{align} 
for appropriately chosen $\alpha_{jl}, \theta_{jl} \in \mathbb{R}$. $A$ achieve their goal by measuring the rotated quadratures with angles $\theta_{jl}$ and summing their results multiplied by $\alpha_{jl}$. 
Since the same reasoning applies to any linear combination of the $\xi_j\mathrm{s} $, $A$ can perform an arbitrary homodyne measurement of the secret $\rho_{\mathrm{s}}$.
Sampling any quadrature from $\rho_{\mathrm{s}}$ allows $A$ to simulate any protocol needing homodyne measurements of $\rho_{\mathrm{s}}$, from quantum key distribution \cite{Grosshans2003}, to measurement based quantum computing \cite{gu2009quantum}, to, when provided with several copies of $\rho_\mathrm{s}$, tomography or verification \cite{ulysse}.

Moreover, $A$ can physically reconstruct the secret state by applying a Gaussian unitary transformation. Let us call $\bs{\xi}^\mathrm{out} \equiv D\bs{\xi}^A $. Since the secret quadratures are conjugated canonical operators we have \begin{equation}
\left[\xi^\mathrm{out}_j,\xi^\mathrm{out}_l \right] = \left[\xi^\mathrm{s}_j,\xi^\mathrm{s}_l \right] = iJ^{\lr{m}}_{jl} .
\end{equation} Since $S_L$ is symplectic, we also have $\left[ \xi^A_j ,  \xi^A_l \right] = iJ^{\lr{k}}_{jl}$. Using $\bs{\xi}^\mathrm{out} = D \bs{\xi}^A$ leads to \begin{equation}
  \left[ \xi^\mathrm{out}_j ,  \xi^\mathrm{out}_l \right]  = i\lr{DJ^{\lr{k}} D^T}_{jl} =iJ^{\lr{m}}_{jl}.
\end{equation} so the rows of $D$ are vectors from a symplectic basis of $\mathbb{R}^{2k}$ \cite{fasanoMarmi} the span of which has dimension $2m$. 
They can be completed to a symplectic basis of $\mathbb{R}^{2k}$ through a Gram-Schmidt-like 
procedure where the scalar product is replaced by the symplectic product \cite{fasanoMarmi}. Alternatively, the procedure explained in Appendix~\ref{sec:symplectification} can be used, improving on the number of required single-mode squeezers (see below). Let us call $S_D^A$ the symplectic matrix the first $m$ and $(k+1)$st to $(k+m)$th rows of which are the rows of $D$, while the others are constructed by one of the above mentioned procedures. Its action on the vector of $2k$ quadratures of the access party $A$ corresponds to a unitary Gaussian transformation $U_D^A$ such that \begin{equation}
  \lr{U_D^A} ^\dagger \bs{\xi} ^A U_D^A = S_D^A \bs{\xi} ^A.
\end{equation} 
By construction, the first $m$ position and momentum entries of $S_D^A\bs{\xi}^A$ correspond to $\bs{\xi}^\mathrm{out}$, so if $A$ apply the physical evolution corresponding to $U_D^A$ and $S_D^A$, they end up with $m$ modes in the secret state, apart from finite squeezing contributions.

Note that $S_D^A$ may, and generally does, involve squeezing. 
However, remarkably, the procedure detailed in Appendix~\ref{sec:symplectification} always allows one to construct $S_D^A$ involving a passive interferometer acting on the $k$ modes of $A$, $2m$ 
independent single-mode squeezers and a final passive interferometer. %
For $m = 1$ (single-mode secrets), the number of squeezers can be further reduced to one per access party by replacing the second one with a homodyne measurement followed by an optical displacement depending on the measurement result. Note that the number of squeezers per access party in the decoding does not scale with the number of players.  This result generalizes the result of \cite{tyc2} to all passive interferometers, including the ones mixing positions and momenta, and to secrets of any size. The generalization beyond orthogonal transformations of the position operators is essential for the result stated in the next section.

\prlsection{Almost any interferometer can be used for QSS \label{sec:haar}} We can now formalize our main result: the encoding and decoding schemes outlined in the previous sections define a secret sharing scheme for almost all passive interferometers $S_L$, in the sense of the Haar measure, that is the constant measure on $L\lr{n}$. In other words, if $S_L$ is drawn uniformly at random from all possible interferometers on $n$ modes, any group of $k$ or more players will almost surely be able decode a secret state of $m$-modes, provided $k\geq m + \lceil\frac{n}{2} \rceil$. A sketch of the proof, detailed in Appendix~\ref{app:proofHaar}, follows.

Let $\mathcal{B}$ be the set of matrices that \emph{cannot} be used for secret sharing. 
For $S_L \in L\lr{n}$ to be in $\mathcal{B}$, $\det\lr{ M \ H^A } = 0$ for at least one access party $A$, 
which we denote $S_L \in \mathcal{B}^A$. 
Because of positivity and countable additivity, we have for the Haar measure of $\mathcal{B}$, $\mu_H\lr{\mathcal{B}}\leq  \sum_A \mu_H\lr{\mathcal{B}^A} $ and we just need to show that each $\mathcal{B}^A$ has zero measure. Each of them is defined as the zero set of a polynomial function of the coefficients of $S_L$ (the determinant of a submatrix), which, regarding $U\lr{n}$ as a manifold, identifies a lower dimensional set, which has zero measure~\cite{knapp2013lie}. In other words, since $L\lr{n}$ is a Lie group of dimension $n^2$, it can be parametrized by $n^2$ real variables defined in an appropriate region $\mathcal{E}\subset \mathbb{R}^{n^2}$. The entries of $S_L$ can be written as polynomials of trigonometric functions of $n^2$ angles $\bs{\lambda}$, so the $\det\lr{ M \ H^A }$ is a real analytic function~\cite{rudin1964principles,krantz2002primer} of $\bs{\lambda}$, whose zero set has necessarily null measure on $\mathcal{E}$~\cite{krantz2002primer,rusRealAnalFun}. 
Therefore $\mathcal{B}$ has zero Haar measure in $L\lr{n}$. Up to a normalization factor, the Haar measure can be seen as a uniform probability distribution over the unitary group. It follows that if a unitary matrix $\bar{U}$ is chosen uniformly at random, the probability that $\bar{U} \in \mathcal{B}$ is zero. 

Note that the \emph{uniformity} property of the Haar measure is not required for the proof: we rather need it to be equivalent to Lebesgue measure on the domain $\mathcal{E}$ of Euler angles, that is if a set has zero measure in $\mathcal{E}$, then it also has zero Haar measure. Our results thus readily apply to any measure that does not assign positive measure to a set of unitaries (interferometers) of zero Haar measure. Moreover, it is not necessary to be able to achieve all possible interferometers in order to find good ones for QSS. To fix the ideas, let us suppose that some experimental setup has continuous parameters $\bs{u}$ that can be adjusted to apply one of a set of interferometers $U\lr{\bs{u}}$.  If $U\lr{\bs{u}}$ spans a set of non-zero Haar measure when $\bs{u}$ is varied, then almost all configurations will lead to a good encoding for QSS according to our definition.

\begin{figure*}
\centering
\subfloat[$m=1$, $n=2$, Haar random]{\includegraphics[width=3.1in]{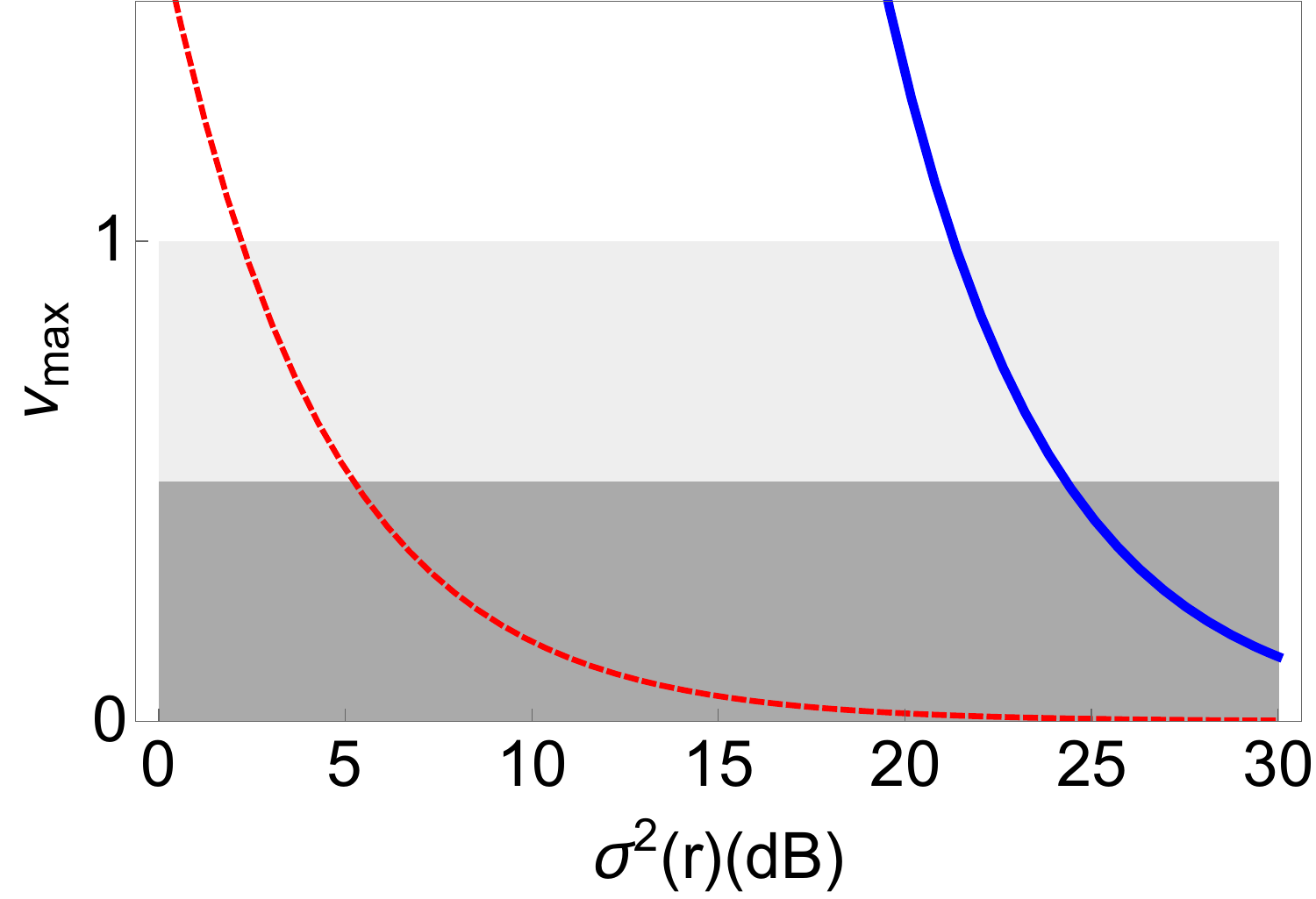}\includegraphics[width=3.1in]{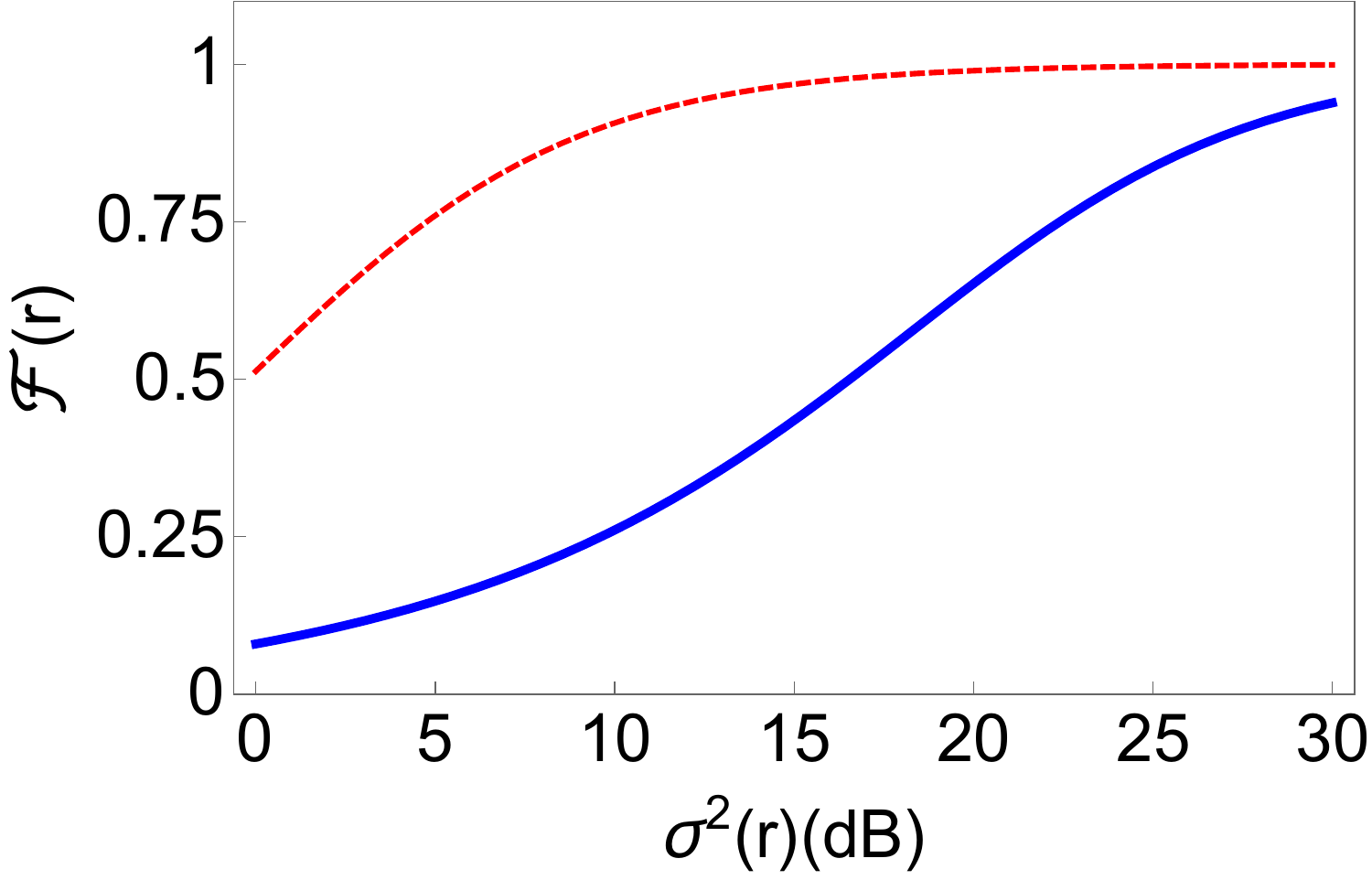} \label{fig:m1n2bis}} \\
\subfloat[$m=1$, $n=2$, best of $10^3$ Haar random matrices]{\includegraphics[width=3.1in]{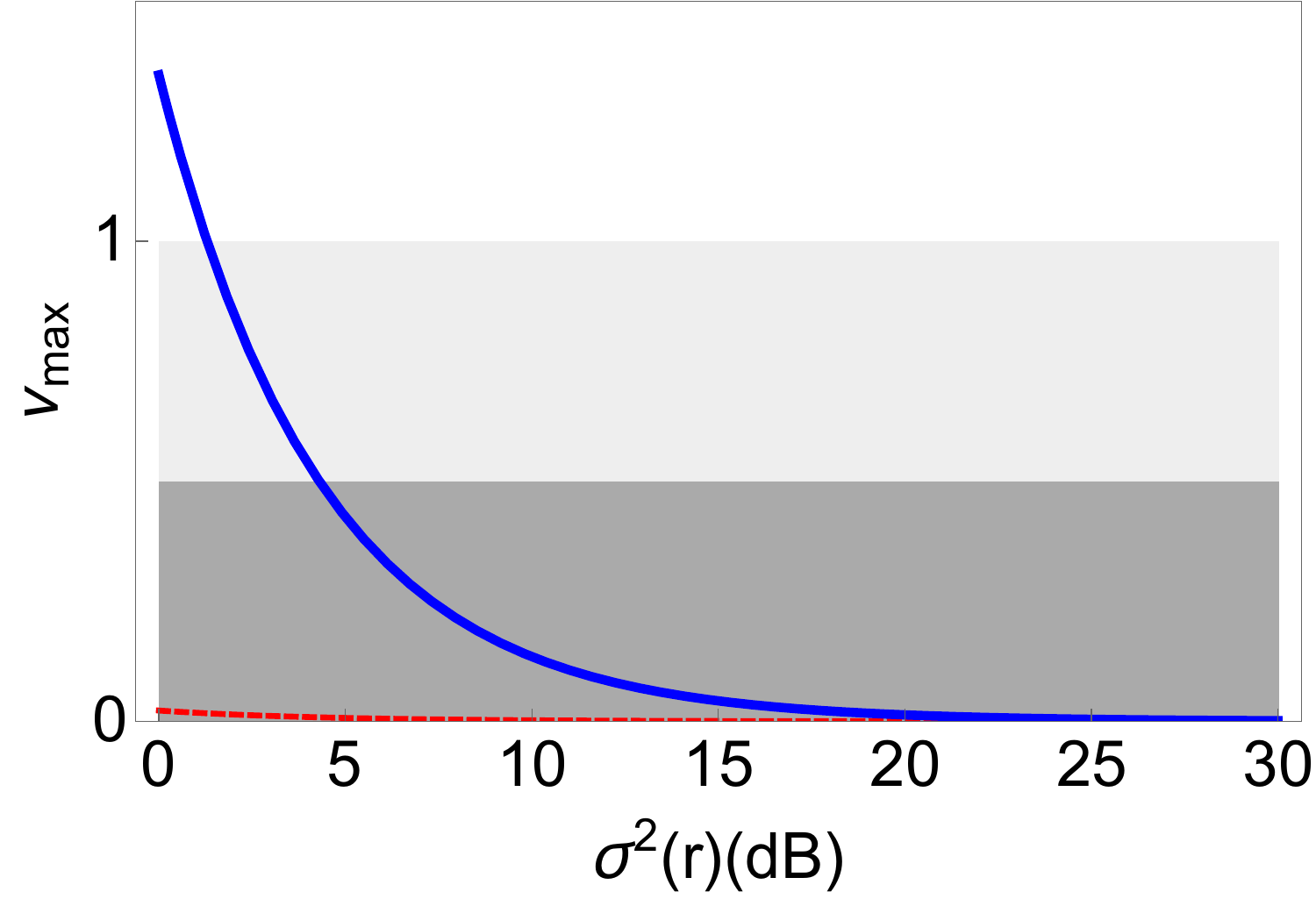}\includegraphics[width=3.1in]{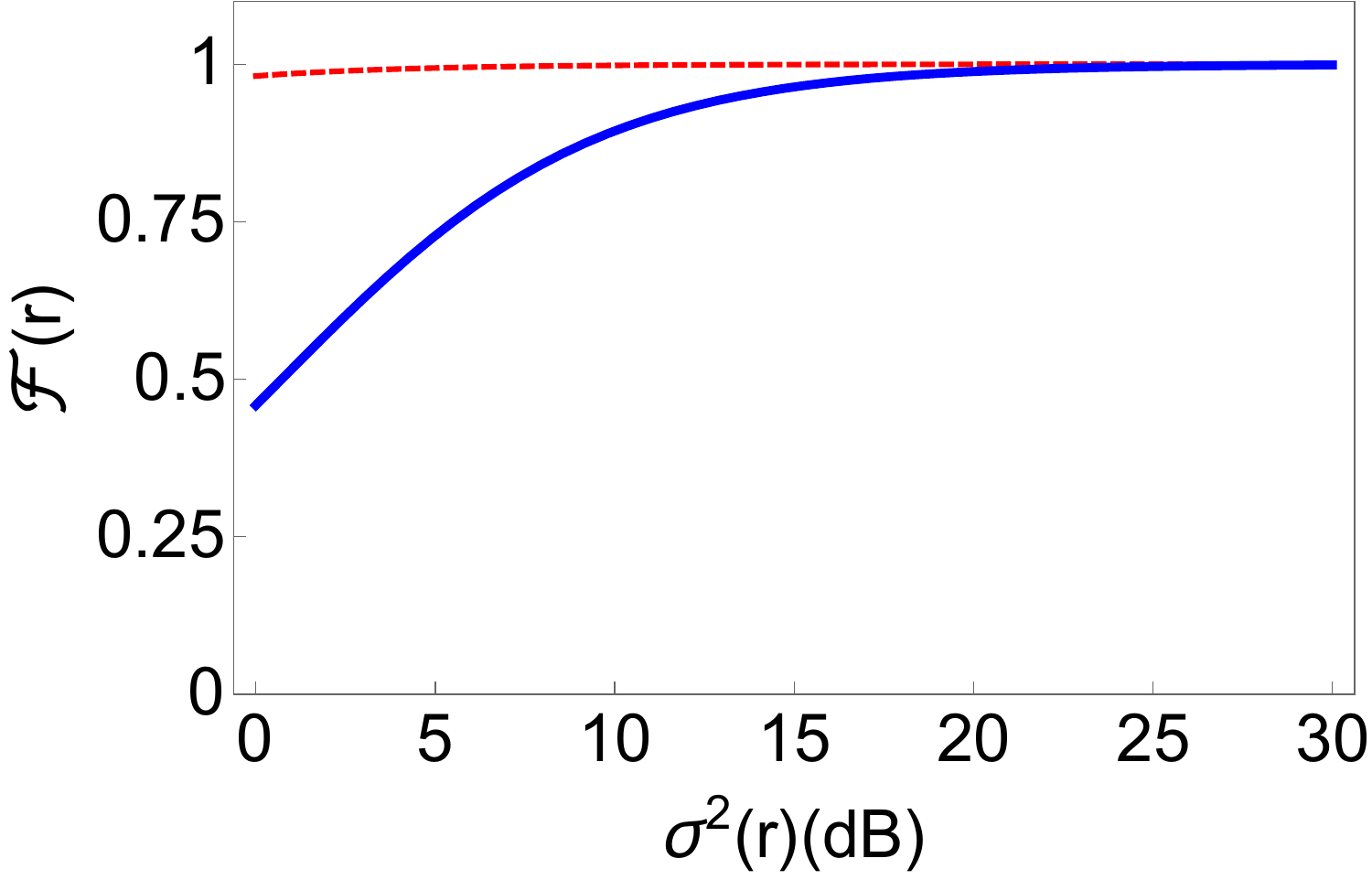} \label{fig:m1n2}} \\
\subfloat[$m=1$, $n=4$, best of $10^3$ Haar random matrices]{\includegraphics[width=3.1in]{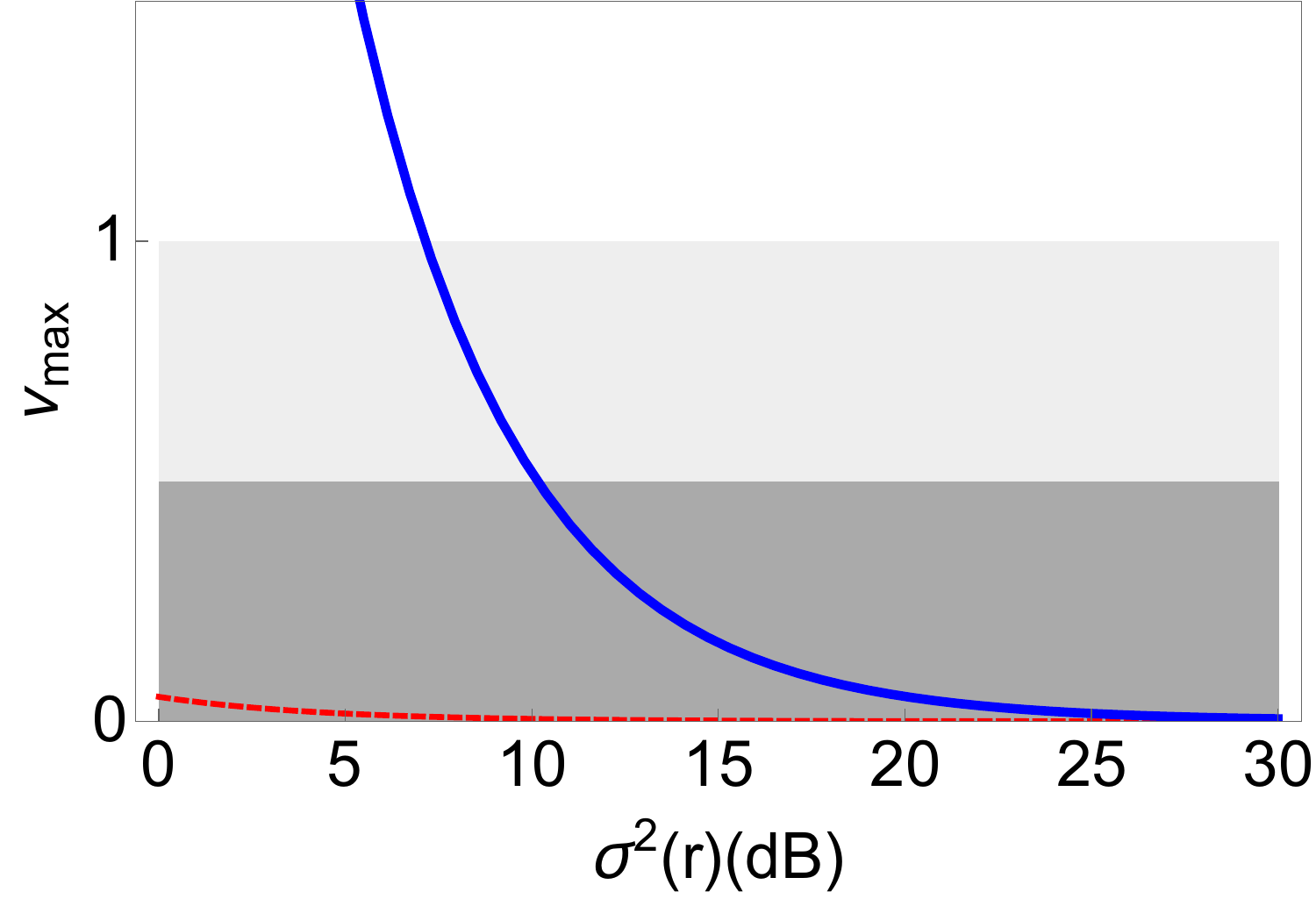}\label{fig:m1n4}}%
\subfloat[$m=2$, $n=2$, best of $10^3$ Haar random matrices]{\includegraphics[width=3.1in]{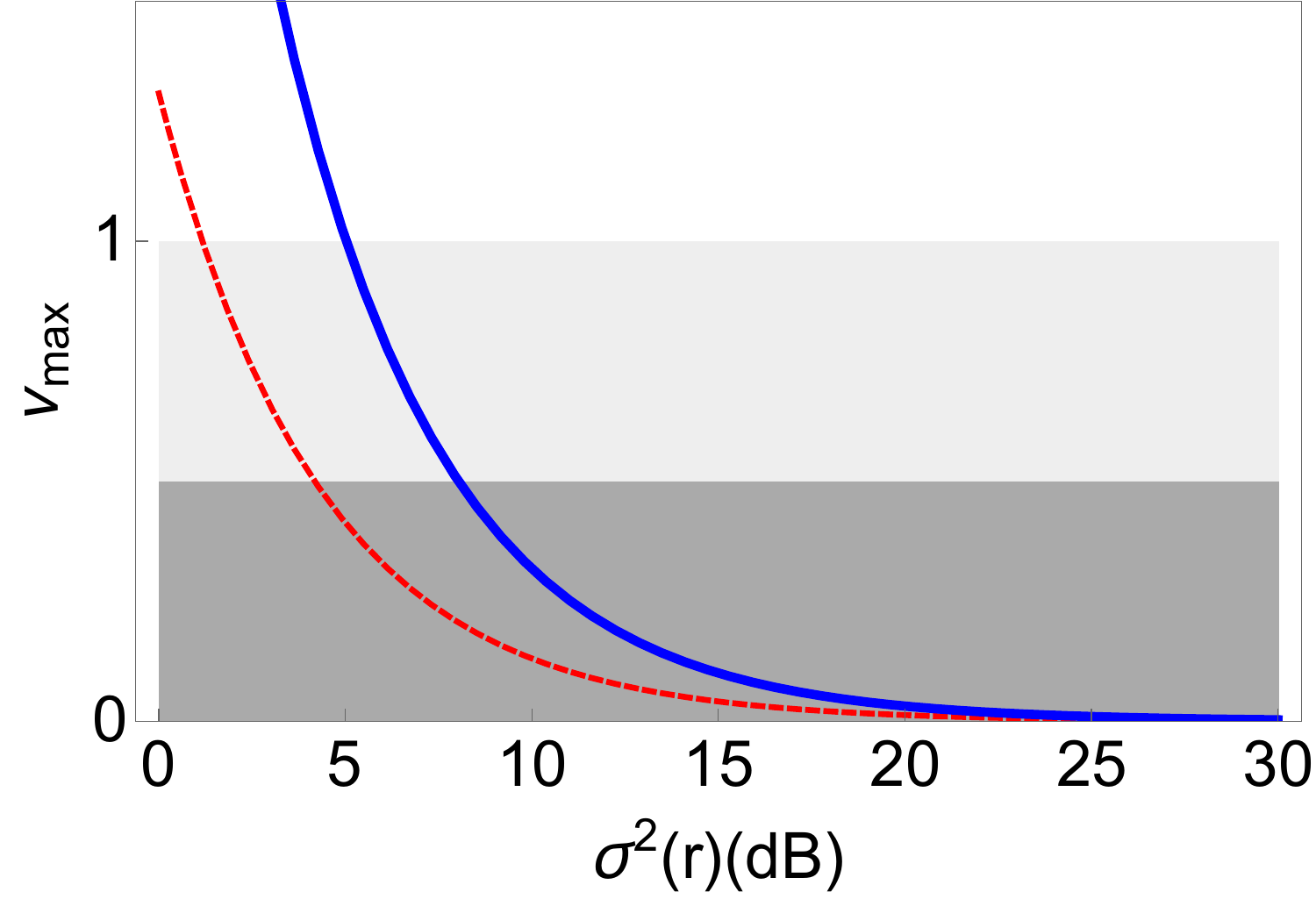}\label{fig:m2n2}}
\caption{Quality of the reconstructed state for several interferometers. The blue, solid (red, dashed) lines correspond to the access party whose decoding results in the worst (best) reconstruction of the secret. In the plots of $\nu_\mathrm{max}$, the upper, white region $\nu_\mathrm{max}>1$ corresponds to the encoding-decoding channel being entanglement-breaking, while the dark gray region $\nu<0.5$ corresponds to the access party having the best possible copy of the secret allowed by optimal cloning. The plots \protect\subref{fig:m1n2bis} and \protect\subref{fig:m1n2} were obtained for two different interferometers and assuming two out of three players are trying to reconstruct a single-mode secret. The plot in \protect\subref{fig:m1n4} is for three out of five players and a single-mode secret, while \protect\subref{fig:m2n2} is for three out of four players trying to reconstruct a two-modes state (the scheme in Fig.~\ref{fig:encoding} ). The matrices representing the corresponding interferometers are  reported in Appendix~\ref{app:matrices}.} 
\label{fig:sqzEff} 
\end{figure*}

\prlsection{Quality of the state reconstructed by authorized sets \label{sec:stuff}} Since both encoding and decoding by any access party require Gaussian resources only, the overall process defines a Gaussian channel~\cite{gaussiInfoRev,eisertChannels}. More specifically, as discussed in Appendix~\ref{sec:SupplgenIn}, the Wigner function of the output state is the one of the secret state convoluted with a Gaussian filter that depends on the initial squeezing and on the interferometer $S_L$. Such channels are sometimes referred to as \emph{(additive) classical noise channels}~\cite{eisertChannels}. In the ideal case where infinitely squeezed states are used $\Delta p_j^\mathrm{sqz} = 0$, the channel coincides with the identity channel. For finite squeezing, the protocol introduces Gaussian noise that becomes smaller as squeezing is increased. We can characterize the quality of the reconstructed state in the realistic imperfect case by relating the amount of input squeezing to the fidelity between the reconstructed state and the secret. 
In particular, suppose for simplicity that the secret is a single-mode coherent state, and all the input squeezed states have the same squeezing $\Delta ^2 p^\mathrm{sqz}_j = e^{-2r}/2 \equiv \sigma^2\lr{r}$. The fidelity of the reconstructed state can then be expressed as (see Appendix~\ref{sec:cohInFid}) \begin{equation}
\mathcal{F}^A\lr{r} =  1/ \sqrt{ 1 + \sigma^2\lr{r} \eta + \sigma^4\lr{r} \zeta } \label{eq:fideMain}
\end{equation} where $\eta = \mathrm{Tr}\lr{BB^T}$, $\zeta = \det\lr{BB^T}$, and $B$ is defined in Eq.~(\ref{eq:simpleDecoded}). Clearly $\mathcal{F}^A\lr{r}\to 1$ for $r\to \infty$. The same holds for any input state, although the expression of the fidelity is generally not as simple.
Another possible way to assess the noise added by the encoding and decoding procedure by one access party is to compute the maximum eigenvalue $\nu_\mathrm{max}$ of the noise matrix $\mathcal{N} = B \Delta^2 B^T$, with $\Delta = \mathrm{diag}\lr{\sigma_1  ,\ \ldots,\ \sigma_{n} }$. This can be interpreted as the size of the smallest features of the secret Wigner function conserved by the channel~\cite{numax1,braunsteinTeleport}. Smaller structures (\emph{e.g.} regions of negativity) are blurred out by the convolution. The values $1$ and $0.5$ can be taken as a references. For $\nu_\mathrm{max}>1$ the channel is known to be entanglement-breaking \cite{braunsteinTeleport,duan,simon}, whereas for $\nu_\mathrm{max}<0.5$ a generalization of the no-cloning theorem ensures that the corresponding access party holds the best possible copy of the secret state~\cite{cerfCloning,fredCloning}. Some examples are plotted in Fig.~\ref{fig:sqzEff}. The squeezing required to achieve a good reconstruction quality depends on the interferometer used for the encoding and can in general be very large (in the tens of decibels). This can be seen from Fig.~\ref{fig:m1n2bis}, reporting $\nu_\mathrm{max}$ and the fidelity obtained from a randomly chosen interferometer as a function of input squeezing. However, interferometers allowing for good reconstruction with technologically achievable squeezing values \cite{sqz1,sqz2} do not seem to be rare and can be found by simple random sampling. Figs.~\ref{fig:m1n2}, \ref{fig:m1n4} and \ref{fig:m2n2} were for example obtained from the interferometers with the smallest $\nu_\mathrm{max}$ from samples of $10^3$ interferometers chosen according to the Haar measure. In general it seems that the required squeezing increases with the number of modes involved but a more thorough characterization of the dependence of the required squeezing on the encoding interferometer is left for future work. The matrices representing the interferometers used for the plots are reported in Appendix~\ref{app:matrices}.

It is worth noting that the quality of the state reconstructed by authorized parties is not affected by the antisqueezing contributions. This means that the same reconstruction quality can be achieved with non pure squeezed states as long as the noise in one quadrature is sufficiently reduced. This particular type of imperfect squeezed states is common in experimental situations. Optical parametric oscillators provide a notable example where the excess noise in the antisqueezed quadrature is larger than the inverse of the noise in the squeezed quadrature (See for example \cite{jacquard}).

If the scheme is used as an ECC, the results of the present section assess how much squeezing is needed to make the secret state robust to the loss of $n+m-k$ modes. For QSS, additional conditions on the unauthorized parties have to be satisfied. This is discussed in the next section.

\prlsection{Unauthorized sets \label{sec:unsets}} In QSS, unauthorized parties should get no information about the secret. This is not strictly true in any realistic realization of our scheme, as for any finite value of squeezing, all subsets can get some information about the secret. This is inherent to any CV protocol, as was recently discussed in detail in \cite{sandersMISS} for a family of single-mode CV-QSS schemes corresponding to a special case of our scheme. As it turns out, in the multi-mode case the access structure is further complicated by the fact that some subsets of less than $m + \lceil\frac{n}{2} \rceil$ players can access some of the secret quadratures even in the infinite-squeezing limit, while groups smaller than a threshold value $k^*$ are prevented from accessing the secret. To see this, let us fix $n$ and $m$ and choose $k = m + \lceil\frac{n}{2} \rceil $. Consider then a set $Z$ of  $k-l$ players, with $0<l<k$. $Z$ can construct a matrix $M^Z$ analogous to $M^A$ but of smaller size. Eq.(\ref{eq:kernel}) then implies that, for $l<m$, $Z$ can almost always retrieve $2m-2l$ combinations of the secret quadratures free from the antisqueezed contributions. On the other hand, a similar reasoning as that in the Sec.~\ref{sec:haar} shows that this is almost never the case for $l\geq m$: groups of  $k-m$ players or less cannot obtain any linear combinations free of all antisqueezing contributions. This implies that they get no information about the secret for infinite input squeezing, as the antisqueezing contributions add white noise to their quadratures. For $m =1$ our scheme defines then a $\lr{k,n}$ threshold scheme \cite{gottesmanTheorySS}, where the adversary structure is composed of all complements of an authorized set. In the general case, the size of the sets that obtain no information about the secret (for infinite squeezing) depends on the size of the latter: any set of $k$ or more players can reconstruct the full secret and sets of $k^* =  k - m$ or less players are denied all 
information about it. Such schemes are known in the DV literature as ramp schemes \cite{ramp} (note that the same term is used with a different meaning in~\cite{sandersMISS}, where only single-mode secrets are considered and the focus is on the information leakage due to finite squeezing).
The amount of information leaked to the adversaries is also constrained  by the fidelity of the state reconstructed by the access party with the secret, since the fidelity of the states reconstructed by disjoint sets of players is limited by optimal cloning. Notably, as mentioned above, increasing the noise of antisqueezed quadratures at fixed squeezing, the reconstruction by the authorized parties is unaffected, while that of the unauthorized party degrades.

\prlsection{Conclusions\label{sec:conclusions}} %
We have introduced a random coding scheme for sharing multimode bosonic states using Gaussian resources. The possibility of using almost any interferometer gives plenty of room for optimization and implies that potentially any experimental setup producing multi-mode squeezed states can be used for QSS, paving the way to quantum resource sharing across entangled networks with arbitrary topology. In particular, this may have applications for sharing resource states in server-client architectures for optical quantum computing \cite{blindCVMBQC, Marshall2016}, which is an increasingly studied paradigm, due to the difficulty of producing genuinely quantum resources for quantum supremacy \cite{NGResource1, NGResource2}. From the perspective of error correction \cite{cleveSS,marin2013equivalence},we can affirm that a Haar randomly chosen linear interferometer acts as an optimal erasure code, since any code tolerating the loss of a higher number of modes would violate no-cloning. %

\begin{acknowledgments}
F. A. is grateful to Luca A. Ardig\`o and Nicolas Treps for fruitful discussions and to Lorenzo Posani for valuable feedback on earlier versions of the manuscript. We acknowledge financial support from the 
BPI France project 143024 RISQ and the French National Research Agency project ANR-17-CE24-0035 VanQuTe. 
\end{acknowledgments}

%%%%%%%%%%%%%%%%%%%%%%%%%%%%%%%%%%%%%%%%%%%%%%%%%%%%%%%%%%%%%%
%%%%%%%%%%%%%%%%%%%%%%%%%%%%%%%%%%%%%%%%%%%%%%%%%%%%%%%%%%%%%%
% Appendix
\appendix

%----------------------------------------------------------------------------------------------------------------------------

\section{Equivalent condition for invertibility of the matrix $T$ \label{sec:eqTinv}}

The decodability conditions derived in the main text are readily computed once $S_L$ is known but checking whether the matrix $T$ (defined in Eq.~(\ref{eq:T}) of the main text) is invertible requires the explicit calculation of a basis of the kernel of $\lr{M^A}^T$ (see Eq.~(\ref{eq:simpleAP})), which is not very practical. We prove here a condition equivalent to the invertibility of $T$ in the case that $M^A$ has full rank: $\mathrm{rank} \lr{M^A} = n-m$. The condition results in a polynomial equation in the coefficients of $M^A$ and thus does not require computing the kernel of $\lr{M^A}^T$ explicitly. This will be particularly useful for the proof of our main result.

Let us call $V = \mathrm{Ker}\lr{\lr{M^A}^T} \subset \mathbb{R}^{2k}$. If $M^A$ has full rank, then $\mathrm{dim}\lr{V} = 2k-n+m = 2m$, since $M^A$ always has $2k$ rows and $n-m$ columns (we assume $k= m + \lceil\frac{n}{2}\rceil$). Let us denote by $\bs{h}_j = H^A\lr{j}$  the $j$th column of $H^A$ and by $\restr{\bs{h}_j}{V}$ its projection on $V$ (see Eq.~(\ref{eq:simpleAP}) for the definition of $H^A$). Let us introduce a basis of $V$, $\left\{v_1,\ldots,v_{2m}\right\}$. We can assume without loss of generality that these vectors are the rows of the matrix $R$ in the main text. Then \begin{equation} \bs{h}_j = \restr{\bs{h}_j}{V} + \restr{\bs{h}_j}{V^\perp} = \sum_l \lr{\bs{v}_l  \cdot \bs{h}_j }\bs{v}_l + \bs{a}_j =  \sum_l T_{lj} \bs{v}_l + \bs{a}_j 
\end{equation} by definition of $T$,  with $\bs{a}_j = \restr{\bs{h}_j}{V^\perp} $.  Consider now the square matrix \begin{equation}
\lr{ M^A\middle| H^A }  = \lr{ M^A  \middle| \bs{h}_1 \middle| \ldots \middle| \bs{h}_{2m} }  
\end{equation} where the notation specifies the last $2m$ columns. Since the determinant is a multilinear, alternating function of the columns we have  
\begin{widetext} 
%\begin{equation}  
\begin{align} 
\mathrm{det}\lr{ M^A  \ H^A } & = \mathrm{det}\left( M^A  \middle| \bs{h}_1 \middle| \ldots \middle| \bs{h}_{2m} \right)  
 = \mathrm{det}\lr{ M ^A \middle| \sum_{l_1} T_{l_1,1} \bs{v}_{l_1} \middle| \ldots \middle| \sum_{l_{2m}} T_{l_{2m},2m} \bs{v}_{l_{2m}} } \notag\\
 & =\smashoperator{\sum_{l_1,\ldots,l_{2m}}} T_{l_1,1}\ldots T_{l_{2m},2m}  \mathrm{det}\lr{ M ^A\middle|  \bs{v}_{l_1} \middle| \ldots \middle| \bs{v}_{l_{2m}} }
  =\smashoperator{\sum_{l_1,\ldots,l_{2m}}} T_{l_1,1}\ldots T_{l_{2m},2m} \epsilon_{l_1,\ldots,l_{2m}} \mathrm{det}\lr{ M ^A\ |  \bs{v}_{1} \ | \ \ldots \ | \ \bs{v}_{2m} }\notag\\
& =\mathrm{det}\lr{T} \mathrm{det}\lr{ M ^A\middle|  \bs{v}_{1} \middle| \ldots \middle| \bs{v}_{2m} }
\end{align}  
\end{widetext} 
where  $\epsilon_{l_1,\ldots,l_{2m}}$ is the completely antisymmetric tensor. The second line follows from the fact that, since $M^A$ is full rank, $V^\perp = \mathrm{span}\lr{\left\{M^A\lr{j}\right\}}$ (in other words, $V$ is the space of the vectors orthogonal to all the rows of $\lr{M^A}^T$). This means that in particular each $\bs{a}_j$ is a linear combination of the rows of $M^A$ so terms containing any of the $\bs{a}_j$ give zero contribution to the determinant. Since by hypothesis $  \mathrm{det}\lr{ M ^A\ |  \bs{v}_{1} \ | \ \ldots \ | \ \bs{v}_{2m} } \neq 0$, it follows that \begin{equation} \label{eq:detCond}
\mathrm{det}\lr{T} \neq 0 \ \iff\   \mathrm{det}\lr{ M ^A\   H^A }  \neq 0.
\end{equation} Since $M^A$ and $H^A$ are defined in terms of the coefficients of $S_L$ and the determinant is a polynomial function thereof, this is the condition we were looking for.

\section{Extending the matrix D to a symplectic matrix \label{sec:symplectification}}

We outline here an algorithm that can be used to extend the matrix $D$ in Eq.~(\ref{eq:simpleDecoded}) for an access party $A$ to a symplectic operation $S^A_D$ corresponding to a physical, unitary Gaussian operation that the access party can implement to output a subsystem in the secret state (apart from terms vanishing for high enough squeezing). It is instructive to begin detailing the case of a single-mode secret state, $m = 1$. The general case is treated in subsection \ref{sec:MMDec}.

Given a subspace $\mathcal{V} \subseteq \mathbb{R}^{2n} $, we will call \emph{symplectic complement} the linear space \begin{equation}
\mathcal{V} ^J \equiv \left\{w \in  \mathbb{R}^{2n} : \langle \bs{v},\bs{w}\rangle = 0\ \forall\ v \in \mathcal{V}  \right\}.
\end{equation} We will reserve the notation $\mathcal{V}^\perp$ and the phrase \emph{orthogonal complement} to indicate the orthogonal complement with respect to the Euclidean product \begin{equation}
\mathcal{V} ^\perp \equiv \left\{w \in  \mathbb{R}^{2n} : \bs{v}\cdot \bs{w} = 0\ \forall\ v \in \mathcal{V}  \right\}.
\end{equation}

\subsection{Single mode secret state}

Let us start from the rows of the matrix $D$ defined in Eq.~(\ref{eq:simpleDecoded}). For $m = 1$, $D$ only has two rows, which we denote by $\bs{x}$ and $\bs{y}$. By construction we have \begin{equation}\begin{aligned}
\bs{x}\cdot \bs{\xi}^A &= q^\mathrm{out} = q^\mathrm{s} + \sum_j B_{1j} p^\mathrm{sqz}_j \\
\bs{y}\cdot \bs{\xi}^A &= p^\mathrm{out} = p^\mathrm{s} + \sum_j B_{2j} p^\mathrm{sqz}_j
\end{aligned}\end{equation} where the matrix $B$ is also defined in  Eq.~(\ref{eq:simpleDecoded}). Our goal is to find $2k-2$ vectors to add as rows of the matrix $D$ such that the resulting matrix is symplectic. To do so, first define \begin{equation}
\bs{x}_1 = \frac{ \bs{x}}{\norm{\bs{x}}}
\end{equation} and $\bs{y}_1 = -J^{\lr{k}}\bs{x}_1$. The vectors $\bs{x}_1$ and $\bs{y}_1$ are both normalized and their symplectic product is $\langle \bs{x}_1, \bs{y}_1 \rangle =1 $, since $\lr{J^{\lr{k}} }^2 = -\mathbb{I}_{2k}$. Using Eq.~(\ref{eq:symplProdComm}) we see that the operators $q_1 = \bs{x}_1\cdot \bs{\xi}^A$ and $p_1 = \bs{y}_1\cdot \bs{\xi}^A$ have the correct canonical commutator $\left[q_1, p_1 \right] = i$. 

Consider now $\mathcal{V}_1\equiv \mathrm{span} \left\{\bs{x}_1, \bs{y}_1 \right\} \subseteq \mathbb{R}^{2k}$ and a normalized vector $\bs{x}_2\in \mathcal{V}_1^\perp$ , that is \begin{equation}
\bs{x}_2\cdot \bs{x}_1 = \bs{x}_2\cdot \bs{y}_1 = 0; \quad  \norm{\bs{x}_2} = 1.
\end{equation} Since $\lr{J^{\lr{k}} }^2 = -\mathbb{I}_{2k}$, these conditions imply that $\bs{x}_2$ has null symplectic product with both $\bs{x}_1$ and $\bs{y}_1$. Moreover, the vector $\bs{y}_2 \equiv -J^{\lr{k}} \bs{x}_2$ is also normalized, orthogonal to  $\bs{x}_1$, $\bs{x}_2$ and $\bs{y}_1$, has null symplectic product with $\bs{x}_1$ and $\bs{y}_1$ and satisfies $\langle \bs{x}_2, \bs{y}_2 \rangle =1$. This is a consequence of $\mathcal{V}_1^\perp = \mathcal{V}_1^J$ and the fact that each multiplication by $J$ transforms Euclidean scalar products into symplectic ones and vice versa, up to a sign. The argument can be repeated for $\mathcal{V}_2 \equiv \mathrm{span} \left\{\bs{x}_1, \bs{x}_2, \bs{y}_1, \bs{y}_2 \right\} \subseteq \mathbb{R}^{2k}$ and a normalized $\bs{x}_3\in \mathcal{V}_2^\perp $ and so on, until $\mathcal{V}_k^\perp = \left\{ \bs{0} \right\}$. The matrix $O_1 = \lr{\bs{x}_1,\ldots,\bs{x}_k, \bs{y}_1,\ldots,\bs{y_k } }^T$ is orthogonal and symplectic by construction, and corresponds to a linear optics transformation leaving the position of the first mode in the secret position, up to a rescaling. The correct scaling can be obtained applying a single-mode squeezer to the first mode, with symplectic matrix $K_1 = \norm{\bs{x}} \oplus \mathbb{I}_{k-1} \oplus \frac{1}{\norm{\bs{x}}} \oplus \mathbb{I}_{k-1} $. 

We now have to ensure that the first mode's momentum is mapped to the secret momentum. Since the rows of $K_1O_1$ are a basis of $\mathbb{R}^{2k}$, we can write \begin{equation}
\bs{y} = \alpha_1 \bs{x} + \sum_{j=2} ^k \alpha_j \bs{x}_j + \frac{ \beta_1 } {\norm{x}} \bs{y}_1 + \sum_{j=2} ^k \beta_j \bs{y}_j. 
\end{equation} It is easy to check that $\langle \bs{x}, \bs{y} \rangle =1 $ implies $\beta_1 = 1$, so \begin{equation}
\bs{y}_1' \equiv \frac{\bs{y}_1}{\norm{\bs{x}}} = \bs{y} - \alpha_1 \bs{x} - \sum_{j=2} ^k \alpha_j \bs{x}_j - \sum_{j=2} ^k \beta_j \bs{y}_j.
\end{equation} Our goal is achieved if we find a symplectic transformation that maps $\bs{y}_1' \mapsto \bs{y}$ leaving $\bs{x}$ unchanged. This is realized in three steps. First, a shear \cite{gu2009quantum} can be applied on the first mode to remove the $\bs{x}$ term. The transformation corresponds to the Gaussian unitary $\mathrm{exp} \lr{i\alpha_1 \bs{q}_1'^2}$, where $\bs{q}_1'$ is the position operator of the first mode after $K_1O_1$ has been applied. The corresponding symplectic matrix is \begin{equation}
K_S = \lr{ \ba{cc} \mathbb{I}_k & 0_k \\ \ba{c} \alpha_1 \ 0 \ \ldots \ 0 \\ 0 \  \ 0 \ \ldots \ 0  \\ \vdots \\ 0 \  \ 0 \ \ldots \ 0 \ea & \mathbb{I}_k \ea} .
\end{equation} Next, rewrite \begin{equation}
\sum_{j=2} ^k \alpha_j \bs{x}_j + \sum_{j=2} ^k \beta_j \bs{y}_j = \sum_{j=2} ^k \eta_j  \lr{ \cos\theta_j\bs{x}_j - \sin\theta_j \bs{y}_j }
\end{equation} and apply mode-wise rotations (phase-shifts) that map \begin{equation}\begin{aligned}
\bs{x_j} & \mapsto \bs{x}_j' =  \cos\theta_j\bs{x}_j - \sin\theta_j \bs{y}_j \\
\bs{y_j} & \mapsto \bs{y}_j' =  \sin\theta_j\bs{x}_j + \cos\theta_j \bs{y}_j,
\end{aligned}\end{equation} which is a passive transformation corresponding to the symplectic matrix  \begin{equation}
O_2= \lr{ \ba{cccc} 1 & 0 \ \ldots & 0 & 0 \ \ldots  \\ 
					\ba{c}0 \\ \vdots \ea & X_2 & \ba{c}0 \\ \vdots \ea & -Y_2 \\
					 0 & 0 \ \ldots & 1 & 0 \ \ldots  \\ 
					 \ba{c}0 \\ \vdots \ea & Y_2 & \ba{c}0 \\ \vdots \ea & X_2 \\
 \ea} .
\end{equation} with $X_2 = \mathrm{diag} \lr{\cos\theta_2,\ldots, \cos\theta_k}$, $Y_2 = \mathrm{diag} \lr{\sin\theta_2,\ldots, \sin\theta_k}$. Finally, apply $k-1$ controlled-$Z$ operations \cite{gu2009quantum} between the first and each of the other modes, of the form $\exp \lr{i\eta_j q_1' \otimes q_j'   }$ with $ q_j' = \bs{x}_j'\cdot \bs{\xi}^A $, $ q_1'   = \bs{x}\cdot \bs{\xi}^A =q^\mathrm{out} $. Each of these two-modes operations acts as \begin{equation}
e^{-i\eta_j q_1'\otimes q_j'  } \lr{\ba{c} q_1' \\ q_j' \\ p' \\ p_j' \ea}  e^{i\eta_j q_1' \otimes q_j'   } = \lr{\ba{c} q_1' \\ q_j' \\ p' + \eta_j q_j' \\ p_j' + \eta_j q_1' \ea}
\end{equation} where $p' = \bs{y}'\cdot \bs{\xi}^A$. Since $\left[q_j',q_l' \right] = 0$ the $C_Z$ operations can be performed in any order, and the resulting symplectic matrix is \begin{equation}
K_2 =  \lr{ \ba{cc} \mathbb{I}_k & 0_k \\
			\ba{cc} 0 & \eta_2 \ \eta_3 \ \ldots \\ \eta_2 & 0\ 0\ \ldots \\ \vdots & \ \ddots \ea & \mathbb{I}_k \\				
 \ea} .
\end{equation} Reconstruction of the secret state at mode one is then achieved by the sequence of transformations corresponding to \begin{equation}
S^A_D = K_2 O_2 K_S K_1 O_1.
\end{equation} This procedure is not efficient in terms of squeezers, as each controlled-$Z$ requires squeezing, and the overall number of independent squeezers required for the above procedure is only upper bounded by the number of modes: it never exceeds $k$, since we could apply the Bloch-Messiah reduction (also known as Euler decomposition) \cite{braunsteinSqueezing,pramana} to $S^A_D$. 

We can however reduce the number of squeezers by the following strategy. Instead of $K_2$, after $O_2$ one could apply a passive transformation that maps \begin{equation}\bs{x}_2' \mapsto \bs{x}_2 '' \propto \sum_{j=2} ^k \eta_j  \lr{ \cos\theta_j\bs{x}_j - \sin\theta_j \bs{y}_j } = \sum_{j=2} ^k \eta_j  \bs{x}_j'. \end{equation} This is always possible, as it amounts to finding a basis of $\mathbb{R}^{k-1}$ the first element of which is proportional to $\lr{\eta_2,\ldots,\eta_k}^T$. Since the $\bs{x}'_j$s are orthonormal, the proportionality constant is $\tilde{\eta} = \lr{\sum_{j=2}^k \eta_j^2}^{-\frac{1}{2}}$. A symplectic orthogonal transformation for the $\lr{2k-2}$-dimensional space $\mathrm{span}\lr{\bs{x}_2',\ldots,\bs{x}_k',\bs{y}_2',\ldots,\bs{y}_k'}$ is obtained imposing that the vectors $\bs{y}_j'$ undergo the same orthogonal transformation. This results in a passive transformation $O_3$ that only acts nontrivially on the last $k-1$ modes and can be grouped with $O_2$. Defining $\tilde{O}_2 = O_3O_2$ we note that $\tilde{O}_2 K_1 = K_1 \tilde{O}_2 $ since the two transformations act on different sets of modes. Reconstruction is then achieved acting a \emph{single} controlled-$Z$ between the first two modes \begin{equation}
\tilde{K}_2 =  \lr{ \ba{cc} \mathbb{I}_k & 0_k \\
			\ba{cc} 0 & \tilde{\eta}^{-1} \ 0 \ \ldots \\ \tilde{\eta}^{-1} & 0\ 0\ \ldots \\ \vdots & \ \ddots \ea & \mathbb{I}_k \\	
 \ea}.
\end{equation} The whole decoding corresponds then to the symplectic matrix \begin{equation}
\tilde{S}^A_D = \tilde{K_2}\tilde{O}_2 K_S K_1 O_1.
\end{equation} Note that $\tilde{O}_2$ can be commuted through $K_SK_1$ and incorporated in $O_1$ to form a global passive transformation. All the squeezing required for the decoding is contained in $ \tilde{K_2} K_S K_1 $ which acts trivially on all but the first two modes, and hence Bloch-Messiah factorization can be applied to factor it as a passive transformation, followed by two independent single-mode squeezers, followed by another passive transformation. In the end, this would lead to a decomposition \begin{equation}
\tilde{S}^A_D = V_2 \Delta V_1
\end{equation} where $V_1$ is a passive transformation on all $k$ modes, $\Delta$ consists of independent squeezers on the first two modes, and $V_2$ is a passive transformation on the first two modes.  

Finally, we note that the number of squeezers can be further reduced to one by replacing the controlled-$Z$ with a homodyne measurement on the second mode followed by a displacement on the first mode depending on the measurement outcome. Indeed, after $K_SO_2K_1O_1$ has been applied, the position quadrature of the first mode is already $\bs{x}\cdot\bs{\xi}^A$, whereas the momentum operator is $\bs{y}\cdot\bs{\xi}^A- \tilde{\eta}^{-1} q_2 ''$ where $q_2 '' = \bs{x}_2''\cdot\bs{\xi}^A$ is the position quadrature of the second mode. If the latter is measured, e.g. by homodyne detection, the operator $ q_2 '' $ is effectively replaced by a real number $\gamma $ and the transformation $\bs{y}\cdot\bs{\xi}^A - \tilde{\eta}^{-1}\gamma \mapsto \bs{y}\cdot\bs{\xi}^A $ can be achieved by a displacement on the first mode.

\subsection{Multi-mode secret state \label{sec:MMDec}}

In the case of a $m$-modes secret state, the matrix $D$ has $2m$ rows and $2k$ columns, with $k$ the number of players in the access party $A$. Recall that \begin{align}
& DJ^{\lr{k}}D^T = J^{\lr{m}}\\
& D\bs{\xi}^A = \bs{\xi}^\mathrm{s} +  B\bs{p}^\mathrm{sqz}.
\end{align} The goal is again to extend the matrix $D$, adding $2k-2m$ rows, in such a way that the resulting matrix is symplectic and maps the quadratures of the first $m$ modes of $A$ to the secret quadratures, apart from distortions due to finitely-squeezed quadratures. In general the resulting matrix will involve squeezing. We aim at minimizing the number of squeezers. To this end, let us consider a matrix $\tilde{D}$, obtained by a partial extension of $D$, that is, $\tilde{D}$ is obtained adding $2l\leq 2k-2m$ to $D$ in such a way that \begin{equation}
 \tilde{D}J^{\lr{k}} \tilde{D}^T = J^{\lr{m+l}}.
\end{equation} If we manage to write $\tilde{D} = S V$ for some symplectic matrix $S \in \mathrm{Sp}\lr{2 m + 2l , \mathbb{R}}$ and $V$ such that $VV^T = \mathbb{I}_{2 m + 2l }$, $VJ^{\lr{k+l}}V^T = J^{\lr{m+l}}$ then decoding can be completed without adding more squeezers to those contained in $S$, the number of which is necessarily smaller than or equal to $m + l$ (as can be seen applying Bloch-Messiah factorization to $S$). In fact,  $VV^T = \mathbb{I}_{2 m + 2l }$ means the rows of $V$ are orthonormal, and since $ \tilde{D}J^{\lr{k}} \tilde{D}^T = J^{\lr{m+l}}$, the orthogonal and symplectic complements of $\mathcal{V} \equiv \mathrm{span} \lr{\left\{ \tilde{D}\lr{j} \right\}}$, where $\left\{ \tilde{D}\lr{j} \right\}$ denotes the rows of $\tilde{D}$,  coincide $\mathcal{V}^J = \mathcal{V}^\perp$. We can thus find an orthonormal basis of $\mathcal{V}^J $ the elements of which are also orthogonal to each vector in $\mathcal{V}$ by the same procedure used in the previous subsection to construct an orthonormal basis of $\mathcal{V}_1$. 

Let us now show that for a general $D$, at most $2m$ rows have to be added. Indeed $\Gamma = DD^T$ and $\Gamma' = \tilde{D}\tilde{D}^T$ are both symmetric, positive-definite matrices. apart from a possible rescaling, they are covariance matrices corresponding to physical states. Requiring that $\tilde{D} = S V$ and simultaneously $VV^T = \mathbb{I}_{2 m + 2l }$ implies \begin{equation}
\Gamma'= SVV^TS^T = SS^T
\end{equation} which means $\Gamma' $ is proportional to the covariance matrix of a pure state. It follows that $\Gamma'$ is (proportional to) a covariance matrix that purifies $\Gamma$. It is known that $m$ ancillas are sufficient to purify a Gaussian state of $m$ modes \cite{gaussiInfoRev}, so $l \leq m $. Since we have no \emph{a priori} information about the structure of $\Gamma$, $m$ ancillary modes are necessary in the worst case. 

Suppose we compute $S$ and $V$ from the purification of $DD^T$. As anticipated, we can extend $V$ to a symplectic orthogonal matrix $O$ with the same procedure used in the previous subsection. We then extend $S$ to a symplectic matrix  $\tilde{S}$ that acts trivially on all but the first $2m$ modes. We can apply Bloch-Messiah reduction to $\tilde{S}$ and decompose it into a passive transformation that can be absorbed in $O$, a matrix $K$ consisting of $2m$ independent squeezers on the first $2m$ modes, and a final passive transformation $V$ acting on the first $2m$ modes, so in the end $S^A_D$ has the form $S^A_D = VKO$.
 
\section{Effect of finite-squeezing noise on the decoded state}

We first show that for general input states the Wigner function of the reconstructed state can be represented as the Wigner function of the input state convoluted with a Gaussian filter function depending on the input squeezing. We then restrict to Gaussian secrets and derive the expression reported in Eq.~(\ref{eq:fideMain}) of the reconstruction fidelity for single mode, coherent input (secret) states.

\subsection{General input states \label{sec:SupplgenIn}}

We now show that for any input state $\rho_\mathrm{s}$, with Wigner function $W_\mathrm{s}\lr{\bxi}$, not necessarily Gaussian, the Wigner function $W_\mathrm{out}\lr{\bxi}$ of the state reconstructed by an access party is given by a convolution of $W_\mathrm{s}$ with a Gaussian filter function. This function is related to the input squeezing, the encoding $S_L$ and the decoding $S^A_D$ and becomes narrower for larger squeezing, eventually converging to a Dirac delta. In this limit, the convolution outputs exactly the secret Wigner function $W_\mathrm{s}$, meaning that the reconstruction is perfect.

Let us start from Eq.~(\ref{eq:simpleDecoded}), which we recall here for convenience \begin{equation}
 \bs{\xi}^\mathrm{out}   = B \bs{p}^\mathrm{sqz} + \bxi^\mathrm{s}  \label{eq:outQuadB}
\end{equation} where $\bs{\xi}^\mathrm{out}\equiv D \bs{\xi}^{A}$ as in the main text. If the matrix $B$ were the zero matrix, then the outcomes of the measurement of any quadrature of the output state would follow the same probability distribution as if the same measurement had been performed on the input state. It follows that the output Wigner function $W_\mathrm{out}\lr{\bxi} $ would be equal to the input Wigner function $W_\mathrm{s}\lr{\bxi} $. If the matrix $B$ is not zero, the output state is obtained by tracing out all squeezed modes. This amounts to averaging over all possible measurement outcomes for the squeezed quadratures $p^\mathrm{sqz}_j$. By assumption, the input modes are independently squeezed, so each $p^\mathrm{sqz}_j$ will contribute with a random shift distributed according to a Gaussian probability density with zero mean and variance $\sigma^2_j = e^{-2r_j}/2$. Since the map that associates a Wigner function to each density matrix is linear, the output Wigner function is \begin{widetext} 
\begin{equation} \label{eq:convP} \begin{aligned}
W_\mathrm{out}\lr{\bxi} &=  \int \lr{\prod_{j=1}^{n}\ddd y_j \frac{ e^{-\frac{y_j^2}{2\sigma^2_j}}}{\sigma_j\sqrt{2\pi}} } W_\mathrm{in} \lr{ \bxi-B\bs{y}} \\
 &= \frac{1}{\det\Delta\lr{2\pi  }^{\frac{n}{2}} }\int \ddd^n y  \exp\lr{-\frac{1}{2} \bs{y}^T\Delta^{-2} \bs{y} }   W_\mathrm{in} \lr{ \bxi-B\bs{y}}
\end{aligned}\end{equation} \end{widetext} 
with $ \Delta = \mathrm{diag}\lr{\sigma_1  ,\ \ldots,\ \sigma_{n} }$. Eq.~(\ref{eq:convP}) is valid for arbitrary input states. The case of a Gaussian $W_\mathrm{in}$ is discussed in the following.

\subsection{Gaussian input states \label{sec:cohInFid}}

Since the protocol only involves Gaussian (squeezed) ancillary states, Gaussian operations (passive interferometers, squeezers) and Gaussian measurement (homodyne), the procedure of encoding and  then decoding can be described as a Gaussian channel. If the input states are also Gaussian, they are fully specified by the quadratures' mean values $\bxi_0$ and covariance matrix $\Gamma$ \begin{equation} \begin{aligned}
\lr{\bxi_0}_j &=  \langle  \xi_j \rangle \\
\Gamma_{jl} &= \langle  \left\{ \xi_j , \xi_l \right\} \rangle.
\end{aligned}\end{equation}  The action of a Gaussian channel can then be described as~\cite{gaussiInfoRev} \begin{equation}\label{eq:genGaussChRepr} \begin{cases}
\bxi_0 \mapsto \mathcal{T} \bxi_0 + \bs{d} \\
\Gamma \mapsto  \mathcal{T} \Gamma  \mathcal{T}^T + \mathcal{N}
\end{cases}\end{equation} where $\bs{d}\in \mathbb{R}^{2m}$, $\mathcal{T}$ and $\mathcal{N}=\mathcal{N}^T \geq 0$ are $2m\times 2m$ real matrices such that $\mathcal{N} + i J^{\lr{m}} - i  \mathcal{T} J^{\lr{m}}  \mathcal{T}^T \geq 0$.

Let us focus on a single access party $A$. By construction, the quadratures of the reconstructed mode are related to the secret quadratures by Eq.~(\ref{eq:outQuadB}) (Eq.~(\ref{eq:simpleDecoded})). We directly see that $ \mathcal{T} = \mathbb{I}$ and $\bs{d} = \bs{0}$. In order to characterize the channel defined by decoding and reconstruction by $A$ we just need to find $ \mathcal{N}$. This is easily accomplished remembering that the input squeezed and secret modes are not correlated, so that \begin{equation}
 \langle p_j^\mathrm{sqz} p_l^\mathrm{sqz} \rangle =  \langle \xi^\mathrm{s}_a p_l^\mathrm{sqz} \rangle = 0
\end{equation} for any $l,\ a$ and $j\neq l$, whence\begin{equation}
\frac{1}{2} \langle  \left\{ \xi^\mathrm{out}_a, \xi^\mathrm{out}_b \right\} \rangle = \frac{1}{2} \sum _l B_{al} B_{bl} \Delta ^2 p_l^\mathrm{sqz} + \frac{1}{2} \langle  \left\{ \xi^\mathrm{s}_a, \xi^\mathrm{s}_b \right\} \rangle.
\end{equation} Denoting $\Delta^2 = \mathrm{Diag}\lr{\Delta^2  p_1^\mathrm{sqz},\ \ldots,\ \Delta ^2 p_{n}^\mathrm{sqz}}$ and comparing with Eq.~(\ref{eq:genGaussChRepr}) we arrive at \begin{equation}
\mathcal{N} = B \Delta^2 B^T.
\end{equation} For the rest of this section, we restrict for simplicity to the  case where all the modes are squeezed by the same parameter $r$, so that \begin{equation}
\mathcal{N} =\mathcal{N}\lr{r} = \frac{e^{-2r}}{2} BB^T. 
\end{equation}
Suppose furthermore that the secret is a single-mode coherent state $\rho_\mathrm{s} = \Ket{\alpha} \Bra{{\alpha}}$, the covariance matrix of which is proportional to the $2\times 2$ identity matrix $\Gamma = \mathbb{I}_{2}/2$. To compute the Fidelity $\mathcal{F}\lr{\alpha,r}$ as a function of the squeezing parameter for an arbitrary input coherent state $\Ket{\alpha}$ we use the fact that for a pure input state, the fidelity reduces to a trace, which is just an overlap integral, in our case between two Gaussian functions, in the Wigner function formalism~\cite{leonhardt1997measuring} \begin{equation}\begin{aligned}
\mathcal{F}\lr{\alpha,r} &= \Bra{\alpha} \rho^\mathrm{out}\lr{r} \Ket{\alpha} \\ &= 2\pi \int \ddd q\ \ddd p\ W_\alpha \lr{q, p } W_\mathrm{out}^{\lr{r}} \lr{q, p }.
\end{aligned}\end{equation} The Wigner functions of the two states are given by \begin{widetext}\begin{align}
W_\alpha \lr{\bxi} &= \frac{1}{\pi} \mathrm{exp} \left\{  -\lr{\bs{\xi} - \bs{\xi}_0 }^T \lr{\bs{\xi} - \bs{\xi}_0 } \right\} \\
W_\mathrm{out}^{\lr{r}} \lr{\bxi} &=  \frac{\mathrm{det}\lr{\mathbb{I} + 2 \mathcal{N}\lr{r}}^{-\frac{1}{2}}}{\pi} \mathrm{exp} \left\{  -\lr{\bs{\xi} - \bs{\xi}_0 }^T \lr{\mathbb{I} + 2 \mathcal{N}\lr{r}}^{-1}  \lr{\bs{\xi} - \bs{\xi}_0 } \right\} 
\end{align} so that $\mathcal{F}\lr{\alpha,r}$ reduces to the Gaussian integral  \begin{equation}
\mathcal{F}\lr{\alpha,r} = \frac{2}{\pi} \mathrm{det}\lr{\mathbb{I} + 2 \mathcal{N}\lr{r}}^{-\frac{1}{2}} \int \ddd ^2 \bs{\xi} \mathrm{exp} \left\{  -\bs{\xi} ^T \left[ \mathbb{I} + \lr{\mathbb{I} + 2 \mathcal{N}\lr{r}}^{-1} \right] \bs{\xi}\right\} 
\end{equation}  \end{widetext}

where we used the fact that the integral does not change with the change of variable $\bs{\xi} \to \bs{\xi} + \bs{\xi}_0$. Standard integration techniques lead to \begin{widetext} \begin{equation}\begin{aligned}
\mathcal{F}\lr{\alpha,r} &= \frac{2}{\pi} 2\pi \left[ \mathrm{det}\lr{\mathbb{I} + 2 \mathcal{N}\lr{r}} \right] ^{-\frac{1}{2}} \left\{ \mathrm{det}\left[ 2\lr{ \mathbb{I} + \lr{\mathbb{I} + 2 \mathcal{N}\lr{r}}^{-1} } \right]\right\}^{-\frac{1}{2}} \\
&= 4 \left[  \mathrm{det}\lr{\mathbb{I} + 2 \mathcal{N}\lr{r}} \right]^{-\frac{1}{2}} \frac{1}{2} \left\{ \mathrm{det}\left[ \mathbb{I} + \lr{\mathbb{I} + 2 \mathcal{N}\lr{r}} ^{-1} \right]\right\} ^{-\frac{1}{2}} \\
&= 2  \mathrm{det} \left\{  \mathbb{I} + 2 \mathcal{N}\lr{r} + \mathbb{I} \right\} ^{-\frac{1}{2}} = \frac{1}{\sqrt{ \mathrm{det}\lr{\mathbb{I} + \mathcal{N}\lr{r}}}} \label{eq:fideN}
\end{aligned} \end{equation} \end{widetext} where we used the fact that for a real number $x$ and an $l\times l$ matrix $M$ one has  $\mathrm{det}\lr{x M} = x^l \mathrm{det}\lr{ M}$ and Binet's formula to go from the second to the third line.

Now, by construction $\mathcal{N}\lr{r} = \mathcal{N}\lr{r}^T \geq 0 $ and $\mathcal{N}\lr{r}\to 0$ for $r\to \infty$, so $\mathcal{F}\lr{\alpha,r} \to 1$ for $r\to \infty$. Moreover, we can derive the simple expression of Eq.~(\ref{eq:fideMain}) by noting that there always exists an orthogonal matrix $O$ such that $OBB^TO^T = \mathrm{diag}\lr{\mu, \nu}$ and since the determinant and the trace are invariant under orthogonal transformations we have, after some algebra, \begin{equation} \begin{aligned}
\mathrm{det}\lr{\mathbb{I} + \mathcal{N}\lr{r}} &= 1 + \frac{e^{-2r}}{2} \lr{\mu + \nu} + \frac{e^{-4r}}{4} \mu\nu \\ &= 1 + \frac{e^{-2r}}{2} \mathrm{Tr}\lr{BB^T} + \frac{e^{-4r}}{4} \mathrm{det}\lr{BB^T}
\end{aligned}\end{equation} which plugged into Eq.~(\ref{eq:fideN}) leads to the desired expression.

\section{Proof that the Haar measure of $\mathcal{B}$ is zero}

\label{app:proofHaar}

We outline here a proof of the fact that the set $\mathcal{B}$ of matrices that cannot be used for secret sharing has zero Haar measure. We first note that integration with respect to the Haar measure of a function defined on $U\lr{n}$ can be written as an ordinary integral over some real variables. We then recall a parametrization of $U\lr{n}$ providing a realization of said variables. Finally, we conclude the proof linking the decodability conditions to the zero set of real analytic functions.

\subsection{Haar measure in terms of real variables}

Although the treatment could apply to more general situations, let us consider directly the case of $U\lr{n}$. Since the unitary group is a real Lie group of dimension $n^2$, we can find an atlas, that is, a family of pairs  $\left\{\lr{V_j,\gamma_j}\right\}$ such that the open sets $V_j \subseteq U\lr{n} $ cover $U\lr{n}$ and each map $\gamma_j :V_j  \to \mathbb{R}^{n^2}$ is a homeomorphism. For any function $f$ defined on $U\lr{n}$ we can define a function $g$ on $\mathcal{E } = \bigcup_j \gamma_j\lr{V_j}\subseteq \mathbb{R}^{n^2}$ as \begin{equation}
g\lr{\bs{\lambda}} = f\lr{\gamma_j^{-1}\lr{\bs{\lambda}}}
\end{equation} for all $\bs{\lambda}\in \mathcal{E} \cap  \gamma_j \lr{ V_j}$. Using the theorem of change of variable, we can then find real valued functions $\Delta_j\lr{\bs{\lambda}}$ such that we can  write any integral with respect to the Haar measure, which we denote by $\ddd \mu ^H$, as an integral over a region of $\mathbb{R}^{n^2}$ \begin{equation}
\int _{V_j} f\lr{\alpha} \ddd\mu^H\lr{\alpha} = \int_{\gamma_j \lr{V_j}} f\lr{\gamma^{-1}_j \lr{\bs{\lambda}}} \Delta_j\lr{\bs{\lambda}} \ddd^{n^2} \lambda.
\end{equation} The integral over the whole unitary group can be defined appropriately gluing together the charts $\left\{\lr{V_j,\gamma_j}\right\}$ \cite{knapp2013lie}. 

\subsection{Parametrization of $U\lr{n}$}

Instead of an atlas, we consider here a single chart which covers \emph{almost all} of $U\lr{n}$ (we will not prove this). This is sufficient for our goals.

In particular, we will consider the parametrization in terms of Euler angles that was used in \cite{cue} to numerically generate Haar distributed unitary matrices. It relies on the fact that any unitary matrix $\alpha \in U\lr{n}$ can be obtained as the composition of rotations in two-dimensional subspaces. Each elementary rotation is represented by a $n\times n $ matrix $E^{\lr{j,k}}$ the entries of which are all zero except for \begin{equation}\begin{aligned}
E^{\lr{j,k}}_{ll} &= 1 \quad \mathrm{for\ } l=1,\ 2,\ \ldots,\ n-1 \quad l\neq j,\ k \\
E^{\lr{j,k}}_{jj} &= \cos \lr{\phi_{jk}}e^{i\psi_{jk}} \\
E^{\lr{j,k}}_{jk} &= \sin \lr{\phi_{jk}}e^{i\chi_{jk}} \\
E^{\lr{j,k}}_{kj} &= -\sin \lr{\phi_{jk}}e^{-i\chi_{jk}} \\
E^{\lr{j,k}}_{kk} &= \cos \lr{\phi_{jk}}e^{-i\psi_{jk}}\end{aligned}
\end{equation} From these elementary rotations one can construct the $n-1$ composite rotations \begin{equation}\begin{aligned}
E_1 &= E^{\lr{1,2}}\lr{\phi_{12},\psi_{12},\chi_{1}}\\
E_2 &= E^{\lr{2,3}}\lr{\phi_{23},\psi_{23},0} E^{\lr{1,3}}\lr{\phi_{13},\psi_{13},\chi_{2}} \\
E_3 &= E^{\lr{3,4}}\lr{\phi_{34},\psi_{34},0}E^{\lr{2,4}}\lr{\phi_{24},\psi_{24},0} \\
 &\quad \times E^{\lr{1,4}}\lr{\phi_{14},\psi_{14},\chi_{3}} \\
\vdots & \\
E_{n-1} &= E^{\lr{n-1,n}}\lr{\phi_{n-1,n},\psi_{n-1,n},0} \\ 
& \quad \times E^{\lr{n-2,n}}\lr{\phi_{n-2,n},\psi_{n-2,n},0}\ldots \\
& \quad \times E^{\lr{1,n}}\lr{\phi_{1n},\psi_{1n},\chi_{n-1}}\end{aligned}
\end{equation} and finally any matrix $\alpha\in U\lr{n}$ can be written as \begin{equation}
\alpha = e^{i\eta} E_1 E_2 \ldots E_{n-1}.
\end{equation}  This can be seen as a function defined in the region $\mathcal{E}\subset \mathbb{R}^{n^2}$ that takes $n^2$ angles \begin{equation}\begin{aligned} 0\leq \phi_{jk} < \frac{\pi}{2} &\mathrm{\ for\ } 1\leq j < k\leq n, \\   0\leq \psi_{jk} < 2\pi &\mathrm{\ for\ } 1\leq j < k\leq n , \\  0\leq \chi_l < 2\pi &\mathrm{\ for\ } 1\leq l <  n, \\   0\leq \eta < 2\pi & \end{aligned}
\end{equation}  and outputs a $n\times n$ unitary matrix.  In summary we defined a map $\gamma^{-1}:\mathcal{E}\to V \subset U\lr{n}$ which is one-to-one and the image of which is the whole $U\lr{n}$, except for a set of zero Haar measure. In practice, given any $\bs{\lambda}\in \mathcal{E}$ we can construct the matrix $\alpha = \gamma^{-1}\lr{\bs{\lambda}}$. So for any function $f:U\lr{n}\to \mathbb{R}$ we can define  $g:\mathbb{R}^{n^2}\to \mathbb{R}$ such that $g\lr{\bs{\lambda}} = f\lr{\gamma^{-1}\lr{\bs{\lambda}}}$. If $f$ is measurable with respect to the Haar measure, we can write   \begin{equation}\begin{aligned} \int _{U\lr{n}} f\lr{\alpha} \ddd\mu^H\lr{\alpha} &=\int _{V} f\lr{\alpha} \ddd\mu^H\lr{\alpha} \\ &= \int_{\mathcal{E}} f\lr{\gamma^{-1} \lr{\bs{\lambda}}} \Delta\lr{\bs{\lambda}} \ddd^{n^2} \lambda \end{aligned} \end{equation} with  \begin{equation}
\Delta\lr{\bs{\lambda}} =\frac{1}{\prod \limits _{k=1}^n \mathrm{Vol}\lr{S^{2k-1}}} \lr{ \prod _{1\leq j < k \leq n} \sin ^{2j-1} \lr{\phi _{jk}} }
\end{equation} where $ \mathrm{Vol}\lr{S^{2k-1}} $ is the hypersurface of the $2k-1$ dimensional sphere in $2k$ dimensions~\footnote{For example, for $k=1$, $\mathrm{Vol}\lr{S^{2k-1}} = 2\pi$ is the length of the circle in the plane.}, and \begin{equation}
\ddd ^{n^2} \lambda = \lr{\prod _{1\leq j < k \leq n} \ddd \phi_{jk}  } \lr{\prod _{1\leq j < k \leq n} \ddd \psi_{jk}  } \lr{\prod _{1 \leq  l < n} \ddd \chi_l  } \ddd \eta.
\end{equation} The normalization included in the function $\Delta$ ensures that \begin{equation}
\int _{V} \ddd\mu^H\lr{\alpha} = \int_{\mathcal{E}}  \Delta\lr{\bs{\lambda}} \ddd^{n^2} \lambda= 1.
\end{equation} Now, since $0\leq\Delta\lr{\bs{\lambda}}\leq 1 \ \forall \bs{\lambda}\in \mathcal{E}$ we have \begin{equation}\begin{aligned}
\int _{U\lr{n}} f\lr{\alpha} \ddd\mu^H\lr{\alpha} &= \int_{\mathcal{E}} f\lr{\gamma^{-1} \lr{\bs{\lambda}}} \Delta\lr{\bs{\lambda}} \ddd^{n^2} \lambda \\ & \leq \int_{\mathcal{E}} f\lr{\gamma^{-1} \lr{\bs{\lambda}}} \ddd^{n^2} \lambda. \end{aligned}
\end{equation} What we want to prove is that the integral of the indicator function $\mathbb{I}_\mathcal{B}$ of $\mathcal{B}$ \begin{equation}
\mathbb{I}_{\mathcal{B}}\lr{\alpha} = \begin{cases} 
      1 & \alpha \in \mathcal{B} \\
      0 & \alpha \notin \mathcal{B} \end{cases}
\end{equation} over $U\lr{n}$ with respect to the Haar measure is equal to zero. This will be achieved if we manage to prove that \begin{equation}
\int_{\mathcal{E}} \mathbb{I}_\mathcal{B}\lr{\gamma^{-1} \lr{\bs{\lambda}}} \ddd^{n^2} x = 0
\end{equation} which is equivalent to \begin{equation}
\int_{\gamma\lr{\mathcal{B}}} \ddd^{n^2} \lambda = 0
\end{equation} namely that the image of $\mathcal{B}$ under $\gamma$ has zero measure in $\mathcal{E}$. This is proven in the next section leveraging the fact that through $\gamma^{-1}$ the coefficients of any unitary matrix are written as real analytic functions of the angles. 

\subsection{Real analytic functions}

Our main result then follows from the observation that $\mathcal{B}$ is the union of the zero sets of real analytic functions. Real analytic functions are defined analogously to their complex counterpart: a function $f:\mathbb{R}^N \to \mathbb{R}$ is analytic on an open set $ D $ if it can be represented as the sum of a converging power series  in a neighbourhood of any point $x_0 \in D$ \cite{rudin1964principles}. As in the complex case, a real analytic function is either identically zero, or its zero set has zero measure \cite{rudin1964principles, krantz2002primer} (See also \cite{rusRealAnalFun} for a self-contained proof).

The parametrization of unitary matrices introduced in the previous subsection gives the coefficients of any unitary matrix as a product of trigonometric functions and complex exponentials of the angles. The coefficients of any symplectic orthogonal matrix are real or imaginary parts of a unitary matrix, so they are trigonometric functions of the angles. As it is well known, sine and cosine can always be written as power series. The set of real analytic functions $\mathcal{F}$ is closed under linear combinations with real coefficients and point-wise multiplication~\footnote{If $f\lr{x},\ g\lr{x} \in \mathcal{F}$, then $h\lr{x}= f\lr{x}g\lr{x}\in \mathcal{F}$.}. $\mathcal{F}$ is also closed under quotient as long as the denominator is not equal to zero~\footnote{If $f\lr{x},\ g\lr{x} \in \mathcal{F}$, then the function $h$ defined wherever $f$ and $g$ are both defined and $ g\lr{x} \neq 0 $ as  $h\lr{x}= f\lr{x}/g\lr{x}\in \mathcal{F}$.}. The coefficients $\lr{S_L}_{jl}\lr{\bs{\lambda}}$ are real analytic functions defined on $\mathcal{E}$. For each access party $A$, $\mathrm{det}\lr{ M \ H^A } $ is a polynomial in the entries of $S_L$ and thus defines a real analytic function of the angles in $\mathcal{E}$. It follows that for all $A$, $\gamma^{-1}\lr{\mathcal{B}^A}$ has zero Lebesgue measure on $\mathcal{E}$ This implies that the Haar measure of each $\mathcal{B}^A$ is zero. Positivity and countable additivity of the Haar measure imply $0 \leq \mu_H\lr{\mathcal{B}}\leq  \sum_A \mu_H\lr{\mathcal{B}^A}$, so the Haar measure of $\mathcal{B}$ is also zero. This concludes the proof.

\section{Interferometers for Fig.~\ref{fig:sqzEff} \label{app:matrices}}

We report here the $X$ and $Y$ blocks of the matrices $S_L$ corresponding to the interferometers used for the plots in Fig.~\ref{fig:sqzEff}. apart from that used for Fig.~\ref{fig:m1n2bis}, the matrices were obtained choosing the interferometer that would lead to the lowest value of $\nu_\mathrm{max}$ out of $10^3$ chosen from the Haar measure.
\begin{widetext}

\subsection{Fig.~\ref{fig:m1n2bis}}

\begin{equation}
X =    \left(
\begin{array}{ccc}
 -0.293099 & -0.803506 & -0.311073 \\
 0.128259 & -0.376779 & 0.463209 \\
 -0.633935 & -0.0662967 & 0.145639 \\
\end{array}
\right) \quad
Y = \left(
\begin{array}{ccc}
 0.0921935 & 0.16507 & 0.368724 \\
 0.650109 & -0.23828 & -0.384196 \\
 -0.254222 & 0.352131 & -0.619594 \\
\end{array}
\right)
\end{equation}

\subsection{Fig.~\ref{fig:m1n2}}

\begin{equation}
X =   \left(
\begin{array}{ccc}
 0.596667 & 0.175214 & 0.100266 \\
 0.108915 & 0.458534 & -0.680759 \\
 0.426961 & -0.608681 & -0.134113 \\
\end{array}
\right) \quad
Y =  \left(
\begin{array}{ccc}
 -0.0698255 & 0.405573 & 0.658688 \\
 -0.457902 & 0.174213 & -0.272814 \\
 -0.485058 & -0.440131 & 0.0151496 \\
\end{array}
\right)
\end{equation}

\subsection{Fig.~\ref{fig:m1n4}}

\begin{equation}\begin{aligned}
X &=   \left(
\begin{array}{ccccc}
 0.300365 & 0.29053 & -0.291467 & 0.497589 & -0.0499837 \\
 0.0193436 & -0.0889674 & -0.576899 & 0.216171 & -0.181089 \\
 0.068743 & -0.627185 & 0.0456175 & 0.267772 & 0.488823 \\
 0.313121 & -0.292716 & 0.202423 & -0.254404 & -0.472559 \\
 0.591341 & 0.0132897 & -0.118776 & -0.45464 & 0.0190248 \\
\end{array}
\right)\\
Y &= \left(
\begin{array}{ccccc}
 0.312353 & -0.285854 & 0.469979 & 0.285289 & -0.0937025 \\
 0.0839586 & -0.117954 & -0.320784 & -0.442078 & 0.509978 \\
 0.445916 & -0.00774418 & -0.243163 & 0.0854139 & -0.15446 \\
 0.382669 & 0.26366 & 0.163123 & 0.252382 & 0.425447 \\
 -0.0840343 & -0.513083 & -0.339929 & 0.121405 & -0.16842 \\
\end{array}
\right)
\end{aligned}\end{equation}

\subsection{Fig.~\ref{fig:m2n2}}

\begin{equation}\begin{aligned}
X &=  \left(
\begin{array}{cccc}
 -0.17138 & 0.363352 & 0.220969 & 0.0345219 \\
 0.158628 & -0.268691 & 0.342882 & -0.0159773 \\
 0.478503 & -0.474253 & -0.255255 & 0.12308 \\
 -0.435812 & -0.0371908 & 0.0669927 & -0.343434 \\
\end{array}
\right) \\
Y &= \left(
\begin{array}{cccc}
 -0.529669 & -0.40525 & 0.435797 & 0.392287 \\
 0.460908 & 0.266619 & 0.628541 & 0.325934 \\
 -0.130468 & -0.312016 & -0.235265 & 0.544141 \\
 -0.128694 & 0.486635 & -0.351609 & 0.556099 \\
\end{array}
\right)
\end{aligned}\end{equation}

\end{widetext}

%%%%%%%%%%%%%%%%%%%%%%%%%%%%%%%%%%%%%%%%%%%%%%%%%%%%%%%%%%%%%%
%%%%%%%%%%%%%%%%%%%%%%%%%%%%%%%%%%%%%%%%%%%%%%%%%%%%%%%%%%%%%%
% bibliography

\bibliography{./SSbib}{}

%------------------------------------------------------------------------------------------

%%%%%%%%%%%%%%%%%%%%%%%%%%%%%%%%%%%%%%%%%%%%%%%%%%%%%%%%%%%%%%
%%%%%%%%%%%%%%%%%%%%%%%%%%%%%%%%%%%%%%%%%%%%%%%%%%%%%%%%%%%%%%
% bibliography

\bibliography{./SSbib}{}

%------------------------------------------------------------------------------------------

\end{document}